\documentclass[preprint2,longabstract]{aastex}

\usepackage{multirow}
\usepackage{ulem}
\usepackage{float}

\newcommand{\hyak}{C/1996 B2 Hyakutake}

\shorttitle{Green and red-doublet emissions in comets}
\shortauthors{Bhardwaj and Raghuram}

\begin{document}


\title{\bf Coupled Chemistry-Emission Model for Atomic Oxygen Green and Red-doublet 
Emissions  in  Comet \hyak}


\author{Anil Bhardwaj\altaffilmark{1} and Susarla Raghuram}
\affil{Space Physics Laboratory, 
 Vikram Sarabhai Space Centre, Trivandrum  695022, India.}
\email{ bhardwaj$\_$spl@yahoo.com, anil$\_$bhardwaj@vssc.gov.in,\\ raghuramsusarla@gmail.com}
%

%
\altaffiltext{1}{Corresponding author: Anil Bhardwaj, \\ bhardwaj$\_$spl@yahoo.com,
 \\ Tel.: +91 471 2562330; fax: +91 471 2706535.}


\begin{abstract}
The green (5577 \AA) and red-doublet (6300, 6364 \AA) lines are prompt emissions 
of metastable oxygen atoms in the $^1$S and $^1$D states, respectively, that 
have been observed in several comets. The value of intensity ratio of green to 
red-doublet (G/R ratio) of 0.1 has been used as a benchmark to identify  
the parent molecule of oxygen lines as H$_2$O. 
A coupled chemistry-emission model is developed to study the production and 
loss mechanisms of O($^1$S) and O($^1$D) atoms and the generation of red and 
green lines in the coma of C/1996 B2 Hyakutake. 
 The G/R ratio depends not only on photochemistry, but also on 
the projected area 
observed for cometary coma, which is a function of the dimension of the slit used and 
geocentric distance of the comet. Calculations show 
that the contribution of photodissociation of H$_2$O to the green (red) line emission is 30 to 70\%
(60 to 90\%), while CO$_2$ and CO are the next potential sources 
contributing 25 to 50\% ($<$5\%). The ratio of the photo-production rate of O($^1$S) to 
O($^1$D) would be around 0.03 ($\pm$ 0.01) if H$_2$O is the main 
source of oxygen lines, whereas it is $\sim$0.6 if the parent is CO$_2$. 
Our calculations suggest that the yield of O($^1$S) production in the
photodissociation of H$_2$O cannot be larger than 1\%. The model 
calculated radial brightness profiles of the red and green lines and  G/R ratios 
are in good agreement with the observations made on comet Hyakutake in March 1996.  
\end{abstract}


\keywords{comets: general -- comets: individual (\hyak) -- Molecular processes}


\section{Introduction}
The spectroscopic emissions from dissociative products in cometary coma are often used in estimating  
  production rates of  respective cometary parent species which are sublimating directly 
from the nucleus \citep{Feldman04, Combi04}.  It is a known fact that at smaller ($<$2 AU)
 heliocentric distances, the inner
cometary coma is dominantly composed of  H$_2$O. The infrared
 emissions of H$_2$O molecule are inaccessible from ground  because of 
strong attenuation by the terrestrial atmosphere.
 Since H$_2$O does not show any spectroscopic 
transitions in ultraviolet or visible regions of solar spectrum, one can estimate it's 
abundance indirectly based on the emissions from daughter products, like OH, O and H.
  Thus, tracking  emissions of the dissociative products of
 H$_2$O has  became an important diagnostic tool in estimating the production rate as well as 
in understanding the spatial distribution of H$_2$O  in comets
\citep{Delsemme76,Delsemme79,Fink84,Schultz92,Morgenthaler01,Furusho06}.
 For estimating the density distribution of H$_2$O
 from the emissions of  daughter species,  one has to account 
for photochemistry and associated emission processes.

\begin{figure}[h]  
\centering
 \noindent\includegraphics[width=20pc]{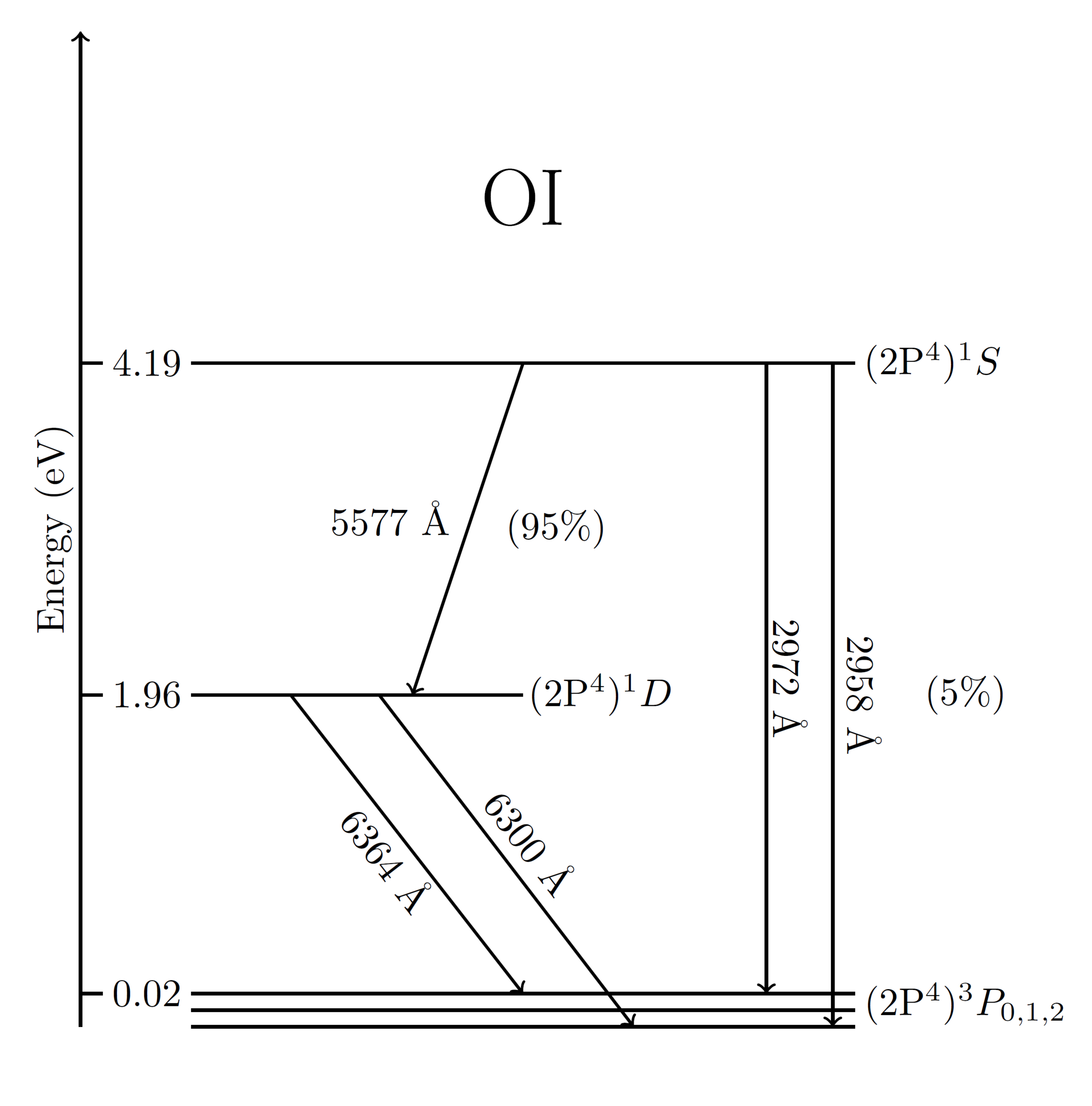}  
\caption{Energy level diagram of atomic oxygen showing different  spectroscopic transitions
related to $^1$S and $^1$D states.}
\label{engyo}
\end{figure}

 The major dissociative channel of H$_2$O is the
 formation of H and OH, but a small fraction  is also possible in 
O($^3$P, $^1$S, $^1$D)  and H$_2$. The radiative decay of  metastable $^1$D and  $^1$S
states of atomic oxygen leads to emissions
at wavelengths 6300, 6364 \AA\ (red doublet) and 5577 \AA\ (green line), respectively. The energy 
levels of atomic oxygen and these forbidden transitions are shown in Figure~\ref{engyo}. Even though
these emissions are accessible from ground-based observatories, most of the times  they
 are contaminated by telluric night sky emissions as well as emissions from other cometary
 species. Doppler shift of these lines, which is 
a function of the relative velocity of comet with respect to the Earth, offers a separation from 
telluric emissions provided a high resolution cometary spectrum is obtained. In most of the cometary
observations it is very difficult to separate the green line in  optical spectrum because 
of the contamination from cometary C$_2$ (1-2) P-branch  band emission. The red line 6300 \AA\ 
emission is also mildly contaminated by the Q-branch emission of NH$_2$ molecule, but in high 
resolution spectrum this can be easily resolved.

Since these atomic oxygen  emissions result due to electronic transitions which are forbidden 
by selection rules, solar radiation cannot populate these excited states directly from 
 the ground state via resonance fluorescence.
The photodissociative excitation and electron
impact excitation of neutral species containing atomic oxygen, and ion-electron dissociative 
recombination of O-bearing ion species, can produce 
these metastable states \citep{Bhardwaj02}. 
 If O($^1$D) is not quenched by ambient cometary species,  then 
photons at wavelengths 6300 and 6364 \AA\ will be emitted in radiative decay  to the ground $^3$P state. 
Only about 5\% of O($^1$S) atoms result in  2972 and 2958 \AA\ emissions via direct radiative transition 
 to the ground $^3$P state of atomic oxygen.
Around 95\% of O($^1$S) decays to the ground state through O($^1$D) by emitting green 
line (cf. Fig. \ref{engyo}).  This implies that if the green line emission is present 
in cometary coma,  the red doublet emission will 
also be present, but the opposite is not always true. The average lifetime of O($^1$D) is relatively 
 small ($\sim$110 s) compared to the lifetime of H$_2$O molecule ($\sim$8 $\times$ 10$^{4}$ s)
 at 1 AU.
  The O($^1$S) also has a very short average  lifetime of about 0.1 s. 
 Due to the short  lifetime of these metastable species, they cannot 
travel larger distances in cometary coma before de-exciting  via radiative transitions. 
Hence, these emissions have been used as diagnostic tools to 
estimate the abundance of H$_2$O in comets \citep{Fink84,Magee90,Morgenthaler01}.
 The intensity of O[I] emissions, in Rayleigh, can be calculated using the following equation
  \citep{Festou81}
\begin{equation}
 I=10^{-6}\tau_p^{-1}\alpha \beta N
\end{equation}
where $\tau_p$ is the lifetime of excited species in seconds, $\alpha$ is the 
yield of photodissociation, $\beta$ is 
the branching ratio, and N is the column density of cometary species in cm$^{-2}$.

In the case of red doublet (6300 and 6364 \AA), since both emissions  arise due to transition 
from the same excited state (2P$^4$ $^1$D) to the ground triplet state (2P$^4$ $^3$P), 
 the intensity ratio of  these two lines should be the same as that of 
branching ratio of  corresponding transitions. Using Einstein transition 
probabilities, \cite{Storey00} calculated the intensity ratio of red doublet 
and suggested that the  intensity of 6300 \AA\  emission 
 would be 3  times stronger than that of 6364 \AA\  emission, and this has been observed  
in several comets also \citep{Spinrad82,Fink84,Morrison97,Cochran01,Capria05,Furusho06,
Capria08,Cochran08}.

 The ratio of intensity of green line  to the sum of intensities of red doublet 
can be calculated as 
\begin{equation}
 \frac{I_{5577}}{I_{6300} + I_{6364}} = \frac{\tau^{-1}_{green} \alpha_{green}N_{green} \beta_{green}}
{\tau^{-1}_{red}\alpha_{red}N_{red}(\beta_{6300+6364})} 
\end{equation}
If the emission intensities of oxygen lines are completely attributed to only photodissociative 
excitation of 
H$_2$O and column densities are assumed almost same for both emissions, then the ratio of intensities of 
green line to red doublet is directly proportional to the ratio of $\tau^{-1}\alpha\beta$. \cite{Festou81}
 reviewed these atomic oxygen emissions in comets. 
Based on the observation of O[I] 2972 \AA\ emission in the IUE spectrograph of  
comet Bradfield (1979X), 
\cite{Festou81} calculated the brightness profiles of red and green emissions. 
 \cite{Festou81} also calculated a
theoretical value for the ratio of the intensity of green line to red doublet 
 (hereafter refer to as the G/R ratio), which has a value of around 0.1
 if H$_2$O is the  source for these O[I] emissions in cometary comae, 
 and it is nearly 1  if the source is CO$_2$ or CO. 
 Observations of  green and red line emissions in several comets have shown that the G/R 
ratio is around 0.1, suggesting that  H$_2$O is the main source  of these O[I] lines. 
However, since no experimental cross section or yield for the production of O($^1$S) from H$_2$O is
available in literature, the  G/R ratio has been questioned by \cite{Huestis06}. 

Generally, the red line is more intense 
than the green line  because the production of O($^1$D) via dissociative excitation of H$_2$O 
is  larger compared to the radiative decay of  O($^1$S).
Since the lifetime of O($^1$D) is larger,  quenching is also a significant loss process
 near the nucleus.  
 So far, the observed G/R ratio  in comets is found to  vary from
 0.022 to 0.3 \citep{Cochran84,Cochran08,Morrison97,Zhang01,Cochran01,Furusho06,
Capria05,Capria08,Capria10}.

 There are several reactions  not involving  H$_2$O which
 can also produce these forbidden oxygen lines \citep{Bhardwaj02}.
Among the O-bearing species, CO$_2$ and CO also have dissociative channels producing O($^1$D) 
and O($^1$S).
However, complex O-bearing molecules (e.g., H$_2$CO, CH$_3$OH, HCOOH) do not produce 
atomic oxygen as a first dissociative product.
Based on the brightness of 6300 \AA\ emission intensity, \cite{Delsemme76}  derived the production 
rate of O($^1$D) in comet Bennett 1970 II and suggested that the abundance of CO$_2$ is more than 
that of H$_2$O. \cite{Delsemme79} estimated the production of O($^1$D) in  
 dissociation of H$_2$O and CO$_2$;  
about 12\% of H$_2$O is dissociated into H$_2$ and O($^1$D), while 67\% of CO$_2$ is 
dissociated into CO and O($^1$D). They suggested  that  a small 
amount of CO$_2$ can contribute
 much more than H$_2$O to the red doublet emission. The model calculations of \cite{Bhardwaj02} 
showed  that the production of O($^1$D) is largely through photodissociative excitation
of H$_2$O while the major loss mechanism in the innermost coma  is quenching by H$_2$O.
 \cite{Cochran01}, 
based  on the observation of width  of red and green lines, 
argued that there must be another potential source  of atomic oxygen in addition 
to H$_2$O, which can produce 
O($^1$S) and O($^1$D).  Observations of the green and red lines in nine comets  
 showed that the green line is wider than the red line \citep{Cochran08}, which could 
be because various parent sources are involved in the  production of O($^1$S). 
 
The model of \cite{Glinski04} showed that the chemistry in the inner coma can produce  
1\%  O$_2$, which can also be a source of red and green lines. 
 \cite{Manfroid07} also argued, based on lightcurves, that forbidden O[I] emissions are probably 
contributed through  dissociation sequence of CO$_2$. 
Recent observation of comet 17P/Holmes  showed that the G/R  ratio 
can be even 0.3, which is the highest reported value so far:  
suggesting that CO$_2$ and CO abundances might be higher at the time of observation \citep{Capria10}.

Considering various arguments based on different observations 
and theoretical  works, we have developed a 
coupled chemistry-emission model  to quantify various  
mechanisms involved in the production of red and green line emissions of atomic oxygen. 
We have calculated the production and loss rates, and the density profiles, of metastable O($^1$D)
 and O($^1$S) atoms from the O-bearing species, like H$_2$O, CO$_2$, and CO,  and also 
from the dissociated 
products OH and O. This model is applied to comet \hyak, which was studied through several observations
in 1996 March \citep{ Biver99, Morrison97, Cochran01, Morgenthaler01,  Combi05, Cochran08}.
 The line-of-sight integrated brightness profiles along cometocentric distances 
 are calculated for 5577 and 6300 \AA\ emissions and 
compared with the observed profiles of \cite{Cochran08}. 
We have also evaluated the role of slit dimension, used in the observation, 
in determining the G/R ratio.
 The aim of this study is to  understand the processes 
 that determine the value  of G/R ratio.

\section{Model}
The neutral parent species  considered 
in this model are H$_2$O, CO$_2$,  and CO. 
We do not consider other significant O-bearing species, like H$_2$CO, CH$_3$OH, since their first 
dissociation does not lead to the formation of atomic oxygen atom; the O atom appears in
 subsequent photodissociation of daughter products, like OH, CO, HCO.
On  1996 March 24, the H$_2$O production rate for comet \hyak\ measured by \cite{Mumma96} was 
1.7 $\times$ 10$^{29}$ s$^{-1}$ .
Based on H Ly-$\alpha$ emission observation, \cite{Combi98} measured H$_2$O production rate as  
2.6 $\times$ 10$^{29}$ s$^{-1}$ on 1996 April 4. Using  molecular radio line 
emissions, \cite{Biver99} derived the production rates of different species at  various
heliocentric distances from 1.6  to 0.3 AU. They found that around 1 AU the relative abundance 
of  CO with respect to H$_2$O is high ($\sim$22\%) in the comet \hyak.

The number density n$_i(r)$ of $i^{th}$ parent species  at a cometocentric
 distance $r$ in the  coma is calculated using the following Haser's formula
\begin{equation}
 n_i(r)=\frac{Q_p}{4\pi v_ir^2} (e^{-\beta_i/r})
 \label{haser}
 \end{equation}
Here $Q_p$ is the total gas production rate of the comet,  $v_i$  and $\beta_i$ are 
the gas expansion velocity (taken as 0.8 km s$^{-1}$, \citeauthor{Biver99} 1999) and the
 scale length ($\beta_{H_2O}$
= 8.2 $\times$ 10$^{4}$ km, $\beta_{CO_2}$ = 5.0 $\times$ 10$^{5}$
km, and $\beta_{CO}$ = 1.4 $\times$ 10$^{6}$ km) of the $i^{th}$ species, respectively.
The Haser model's neutral density distribution has been used in several previous 
studies for deriving the production rate of H$_2$O 
 in comets based on the intensity of  6300 \AA\ emission \citep{Delsemme76,
Delsemme79, Fink84, Morgenthaler01}.  
In our model calculations the  H$_2$O production 
rate on 1996 March 30 is taken as  2.2 $\times$ 10$^{29}$ s$^{-1}$.
 The  abundance of CO relative to H$_2$O is taken 
as 22\%. Since there is no report on the observation of CO$_2$ in the comet Hyakutake,  we 
assumed its abundance as 1\% relative to  H$_2$O. However, we  vary CO$_2$ abundance 
to evaluate its effect on the green and red-doublet emissions.
  The calculations are 
made when the comet \hyak\ was at a heliocentric distance of 0.94 AU and a geocentric distance of 
 0.19 AU on 1996 March 30. The calculated G/R ratio on  
other days of the observation is  also reported.

The number density of OH  produced in dissociation of parent species H$_2$O at a given 
cometocentric distance $r$ is calculated using Haser's two parameter coma model 
\begin{equation}
 n_{OH}(r)= \frac{Q_P}{4\pi   vr^2} \frac{\beta_P}{\beta_R-\beta_P}(e^{-\beta_Pr}-e^{-\beta_Rr})
\end{equation}
Here   $v$ is the average 
velocity of daughter species  taken as 1 km s$^{-1}$, and $\beta_P$ and 
$\beta_R$ are the  destruction scale lengths of the parent (H$_2$O, 8.2 $\times$ 10$^{4}$ km) 
and daughter (OH, 1.32 $\times$ 10$^{5}$ km) species, respectively \citep{Huebner92}.
The solar UV-EUV flux is taken from SOLAR2000 v.2.3.6 (S2K) model of \cite{Tobiska00}
for the day 1996 March 30, which  is shown in Figure~\ref{solflx}. For comparison
the solar flux used by \cite{Huebner92} in calculating O($^1$D) and O($^1$S) production rates 
from various O-bearing species is also presented in the same Figure.

\begin{figure}[h] 
\centering
 \noindent\includegraphics[width=20pc,angle=0]{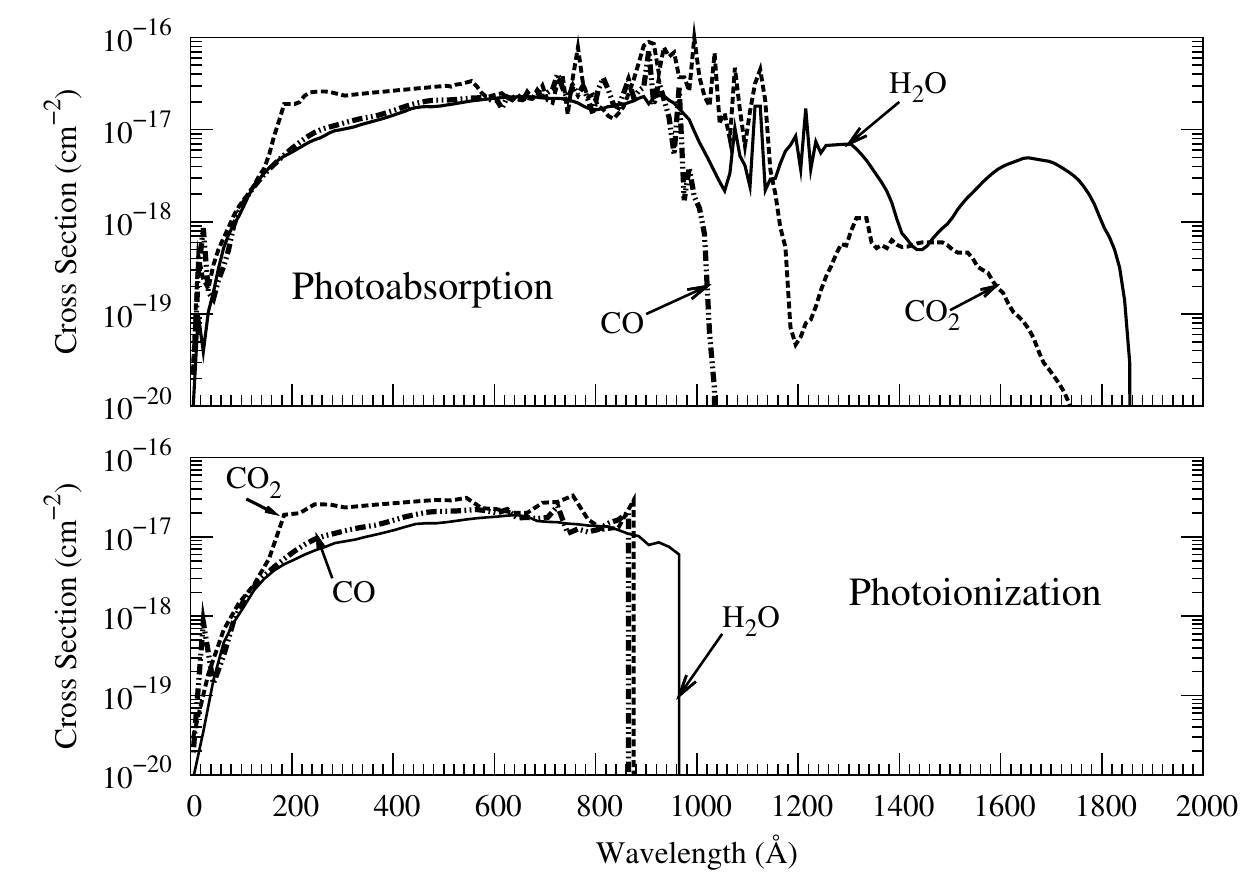} 
\caption{The total photoabsorption cross sections for H$_2$O, CO$_2$ and CO  are shown in the 
top panel and total photoionization 
cross sections  are shown in the bottom panel. The cross sections are taken from \cite{Huebner92}.}
\label{totabcsc}
\end{figure}

The primary photoelectron energy spectrum $Q(E, r, \theta)$
 is calculated by degrading
 solar radiation in the  neutral atmosphere using 
\begin{equation}
Q(E, r, \theta) = \sum_{i}\int_{\lambda}n_i(r)\ \sigma_i^I(\lambda)\ I_{\infty}{(\lambda)}\
exp[-\tau(r,\theta,\lambda)]\ d\lambda 
\label{pheprod}
\end{equation}
where,
\begin{equation}
\tau(r,\theta,\lambda)= \sum_{i}\sigma_i^A(\lambda)\ sec\ \theta \int_r^\infty  n_i(r') dr'  
\end{equation}
Here $\sigma_i^A(\lambda)$  and $\sigma_i^I(\lambda)$ are the absorption and ionization 
cross sections, respectively, of the $i^{th}$ species at the wavelength $\lambda$, 
$n_i(r)$ is its neutral gas density  and
 $\tau(r,\theta,\lambda)$ is optical depth of the medium at the solar zenith angle $\theta$.
 $I_{\infty}(\lambda)$ is the unattenuated solar  flux at the top of atmosphere at wavelength $\lambda$. 
 All calculations are made at solar zenith angle $\theta$ of 0$^0$.
 The total photoabsorption and photoionization cross sections  of H$_2$O, CO$_2$, and CO 
are taken from the compilation of \cite{Huebner92}
 (\url{http://amop.space.swri.edu}),
 and interpolated at 10 \AA\ bins to make them compatible with the S2K solar flux wavelength bins 
for use in our model calculations. 
The total photoabsorption and  photoionization cross sections for H$_2$O, CO$_2$, and CO
 are presented in Figure~\ref{totabcsc}.
The photochemical production rates for ionization and excitation of various species 
are calculated using degraded solar flux and cross sections of 
corresponding processes (discussed in Section~\ref{disso1so1d}) at different 
cometocentric distances.

The primary photoelectrons are degraded in cometary coma to calculate the steady state 
photoelectron flux using 
the Analytical Yield Spectrum (AYS) approach, which is based on the Monte Carlo method
\citep{Singhal91,Bhardwaj93,Bhardwaj99d,Bhardwaj09}. 
The AYS method of degrading electrons in the neutral atmosphere  can be 
explained briefly in the following manner. Monoenergetic electrons incident along Z-axis in an 
infinite medium are degraded in collision-by-collision manner using the Monte Carlo technique.
The energy and position of the primary electron and its secondary or tertiary are recorded at the
instant of an inelastic collision. The total number of inelastic events in the spatial 
and energy bins, after the incident electron and all its secondaries 
and tertiaries have been completely degraded, is used to generate numerical yield
spectra. These yield spectra contain the yield information about the electron degradation process 
and can be employed to calculate the yield for any inelastic event.
The numerical yield spectra generated in this way are in turn
represented analytically, which  contains the information about all possible collisional events 
based on the 
input electron impact cross sections, resulting in the AYS. This yield spectrum can be used 
to calculate the steady state photoelectron flux.
More details of the AYS approach and the method of photoelectron computation are given in 
several  previous  papers 
\citep{Singhal84,Bhardwaj90,Bhardwaj96,Singhal91, Bhardwaj99a,Bhardwaj03,
Bhardwaj99b,Haider05,Bhardwaj09,Bhardwaj11a,Raghuram11}.
The total inelastic electron impact cross sections 
 for H$_2$O are taken from \cite{Jackman77} and \cite{Seng76}, and those for CO$_2$ 
and CO are taken from \cite{Jackman77}.
 The electron impact cross sections for different dissociative ionization states  of H$_2$O 
are taken from \cite{Itikawah2o}, for CO$_2$ from \cite{Bhardwaj09}, and for CO  
from \cite{Mcconkey08}. 
The volume excitation rates for different processes are calculated using steady state 
photoelectron flux and 
electron impact cross sections. The electron temperature required for  
 ion-electron dissociative recombination reactions  is 
taken from \cite{Korosmezey87}.
The detailed description of coupled chemistry-transport model has been given in our earlier papers 
\citep[]{Bhardwaj95,Bhardwaj96,Bhardwaj99a,Bhardwaj02,Haider05,Bhardwaj11}. 
 Various reactions involved in the production and loss  of metastable O($^1$S) and  O($^1$D) atoms 
considered   in our model are listed in Tables~\ref{tab-prlos1s} and~\ref{tab-prlos1d}, 
respectively.

\section{Dissociation of neutral species producing O($^1$S) and O($^1$D)}
\label{disso1so1d}
\subsection{Photodissociation}
\subsubsection{H$_2$O and OH}
\label{phcsch2o}
The  dissociation of H$_2$O molecule starts at 
wavelengths less  than 2424~\AA\ and the primary  products are H and OH.
But the pre-dissociation process mainly starts from 1860 \AA\ \citep{Watanabe53}. The 
threshold wavelength  for the photoionization of H$_2$O is 984 \AA. Hence, solar UV photons in the 
wavelength region 1860 to 984 \AA\ can dissociate H$_2$O and produce different 
daughter products. The threshold wavelengths for the dissociation of H$_2$O resulting in the 
production of  O($^1$S)
 and O($^1$D) are  
1390~\AA\ and 1770~\AA, respectively. Till now,   the 
photo-yield value for the production of O($^1$D) from H$_2$O have been measured 
in only two experiments. 
 \cite{Slanger82} measured the O($^1$D) yield in 
photodissociation of H$_2$O at 1216 \AA, and found  its value to be 10\%. 
  \cite{Mcnesby62} reported a 25\% 
yield for the production of O($^1$D) or O($^1$S) at 1236 \AA\ from H$_2$O.

\begin{figure}[h] 
\centering 
  \noindent\includegraphics[width=20pc,angle=0]{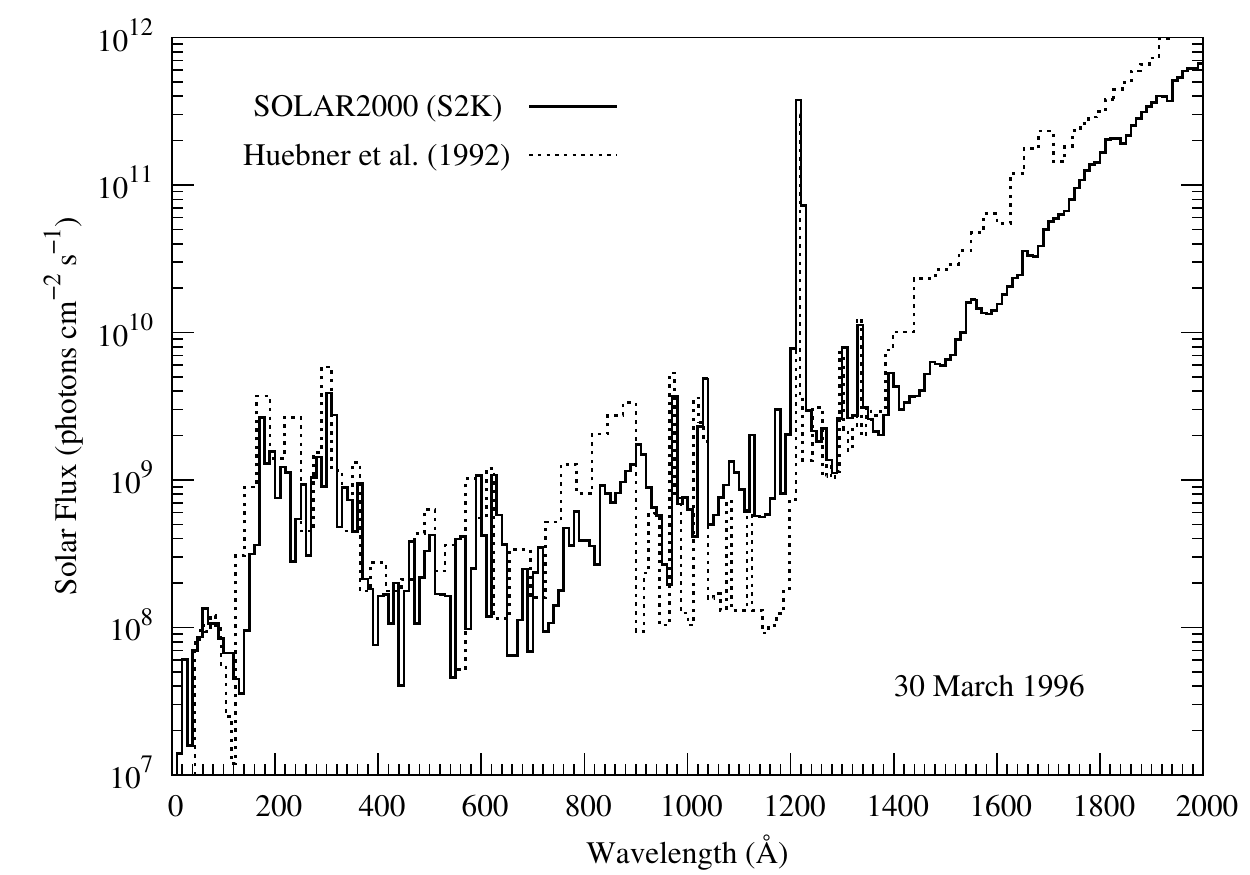}  
\caption{Solar EUV-UV flux  from SOLAR2000 (S2K) model of \cite{Tobiska04} for the day 1996 March 30 
 at heliocentric distance of 1 AU. For comparison the solar flux used by \cite{Huebner92} is also shown.
 At H Lyman $\alpha$ (1216 \AA) the solar flux given by S2K model is higher than that of 
\cite{Huebner92} by a factor 1.24. Significant differences in the two solar fluxes can be noticed in the 
wavelength ranges 800 to 1200 \AA, while above 1400 \AA\ the solar flux of S2K model is smaller than that 
of \cite{Huebner92}. }
\label{solflx}
\end{figure}

\cite{Huebner92} calculated photo production rates for  different excited species produced from 
H$_2$O  using absorption and ionization cross sections compiled  from different experimental  
measurements. 
In our model the cross sections for the production of  O($^1$D) in photodissociation
 of H$_2$O are taken from  \cite{Huebner92}, which were determined based on experiments 
of \cite{Slanger82} and \cite{Mcnesby62}. 
 \cite{Huebner92} assumed that  
 in the 1770 to 1300 \AA\ wavelength region around 25\% of H$_2$O molecules photodissociate 
into H$_2$ and O($^1$D), while between 1300 and 984 \AA\ about 10\%    
of H$_2$O dissociation  produces O($^1$D) (cf. Fig.~\ref{phcsco1d-1}). Below 984 \AA, \cite{Huebner92}
assumed that 33\% of dissociation of H$_2$O leads to the formation of O($^1$D). 
\cite{Festou81a}  discussed various dissociation channels for H$_2$O in the wavelength region
less than 1860 \AA. Solar photons  in the wavelength region 1357 to 1860 \AA\ dissociates around 72\% of 
H$_2$O molecules into ground states of H and OH. But, according to \cite{Stief75}  
approximately 1\% of  H$_2$O molecules are dissociated into H$_2$ and O($^1$D) in this wavelength region.
 The calculated rates for the  O($^1$D) production  from photodissociative excitation 
of H$_2$O by \cite{Huebner92} are 
5.97 $\times$ 10$^{-7}$ s$^{-1}$ and 1.48 $\times$ 10$^{-6}$ s$^{-1}$ for solar quiet
and active conditions, respectively. 
Using the S2K solar EUV-UV flux on 1996 March 30 and cross sections  
 from  \cite{Huebner92} (see Figure~\ref{phcsco1d-1}),
our calculated value is 8 $\times$ 10$^{-7}$ s$^{-1}$ (cf. Table~\ref{tab-prlos1d}), 
 which is a factor of $\sim$1.5 higher  than that of \cite{Huebner92} for solar minimum condition at 1 AU. 
This difference in calculated values is mainly due to the higher (a factor of 1.24) value of solar 
flux  at 1216 \AA\ in S2K model than that  used by \cite{Huebner92}  (cf. Figure~\ref{solflx}).

 No experimentally determined cross sections  for 
 the production of O($^1$S) in photodissociation of H$_2$O are available.
The solar flux  at H Lyman-$\alpha$ (cf. Fig.~\ref{solflx}) is more than an order of magnitude 
larger  than the flux at wavelengths below 1390 \AA, which is the threshold for the  O($^1$S) production 
in dissociation of H$_2$O.
To account for the production of O($^1$S) in photodissociation of H$_2$O, we assumed an 
 yield of 0.5\% at solar H Lyman-$\alpha$ (1216 \AA). However, to assess the impact of this assumption 
on the  green and red 
line emissions we varied the yield between 0 and 1\%.
The calculated photo-rate for the production of O($^1$S) from H$_2$O is
 6.4 $\times$ 10$^{-8}$ s$^{-1}$ at 1 AU assuming 1\% yield at 1216 \AA\ (cf. Table~\ref{tab-prlos1s}).

The primary dissociative product of H$_2$O is OH. The important destruction mechanisms  of OH molecule
 are pre-dissociation through fluorescence process and direct photodissociation. The solar radiation 
shortward of 928 \AA\ can ionize  OH molecule.
The threshold wavelengths for the production of O($^1$D) and O($^1$S) in  photodissociation of 
OH are 1940 and 1477 \AA, respectively. 
 The dissociation channels of OH have been discussed by \cite{Budzien94} and \cite{Dishoeck84}.
 We have used the photo-rates 
given by \cite{Huebner92} for the production  of O($^1$D) and O($^1$S) 
from OH molecule whose values are 6.4 $\times$ 10$^{-7}$ and 
6.7 $\times$ 10$^{-8}$ s$^{-1}$, respectively. These rates are based on dissociation cross sections 
of \cite{Dishoeck84}, which are consistent with the red line observation made by 
wide-field spectrometer
\citep{Morgenthaler07}.

\subsubsection{CO$_2$}
\begin{figure}[h]  
\centering
\noindent\includegraphics[width=20pc,angle=0]{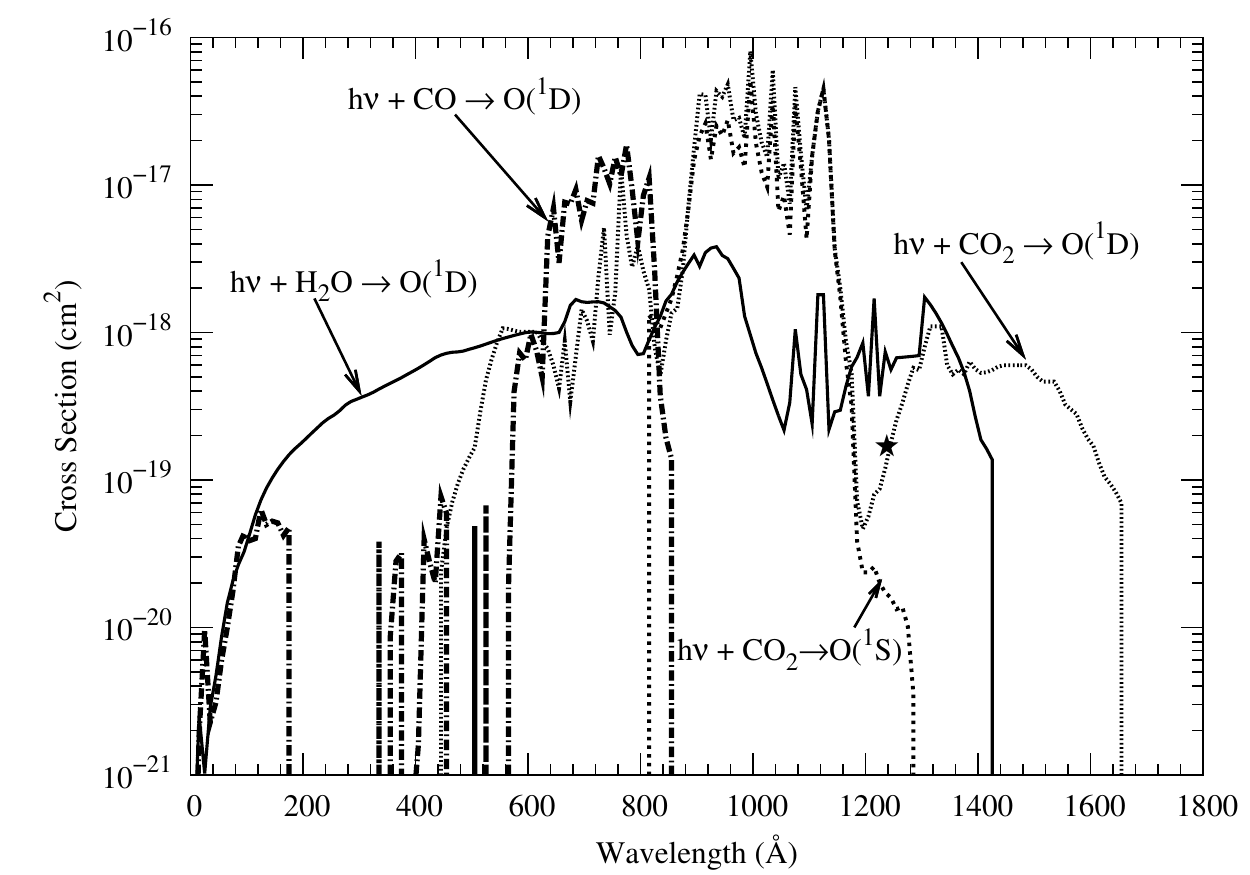}
\caption{Photodissociative excitation cross sections for the production of  O($^1$D)
  from H$_2$O, CO$_2$, and CO. These cross sections are 
taken from \cite{Huebner92}. $\bigstar$ represents the cross section value 
for the production of O($^1$S) from H$_2$O at 1216 \AA\ assuming 1\% yield.}
\label{phcsco1d-1}
\end{figure}

The threshold wavelengths for dissociation of CO$_2$ molecule  
producing  O($^1$D) and O($^1$S) are
1671 \AA\ and  1286 \AA, respectively. As noted by \cite{Huestis06}, the O($^1$D) 
yield in photodissociation of CO$_2$ has never been measured   because of the problem of
 rapid quenching of this metastable state.
  However, experiment by \cite{Kedzierski98} suggested that this dissociation channel can be studied  
in electron impact experiment using solid neon matrix as detector. 
\cite{Huebner92}  estimated the cross section  for O($^1$D) production in 
photodissociative
 excitation of CO$_2$ (see Figure~\ref{phcsco1d-1}),
and obtained photo-rate values of  9.24 $\times$ 10$^{-7}$ and 
1.86 $\times$ 10$^{-6}$ s$^{-1}$ for solar minimum  and maximum conditions, respectively. 
Using S2K solar flux on 1996 March 30 our calculated 
rate for O($^1$D) production in photodissociation of CO$_2$ is 
1.2 $\times$ 10$^{-6}$ s$^{-1}$ at 1 AU, which is 
 higher than the solar minimum  rate of \cite{Huebner92} by a factor of 1.3.
This variation is mainly due to the differences in the solar fluxes (cf.~Figure~\ref{solflx}) 
 in the wavelength region 950 to 1100 \AA\ where the 
photodissociative cross section for the production of O($^1$D)
  maximizes (cf.~Figure~\ref{phcsco1d-1}).

\cite{Lawrence72a} measured the O($^1$S) yield  in photodissociative 
excitation of CO$_2$  from threshold (1286 \AA) to 800 \AA. The yield of 
\cite{Lawrence72a} is different from that measured by \cite{Slanger77}
 in the 1060 to 1175 \AA\ region. However, the yield from both experimental 
measurements closely matches in the 
 1110--1140~\AA\ wavelength region, where the yield is unity. In the experiment of
\cite{Slanger77},  a dip in quantum yield  is observed at 1089 \AA.
\cite{Huestis10} reviewed the  experimental results and suggested the yield for O($^1$S) in 
photodissociation of CO$_2$.
We calculated the  cross section for the O($^1$S) production in photodissociative excitation 
of CO$_2$ (see Figure~\ref{phcsco1d-1}) by multiplying the yield recommended by \cite{Huestis10} with 
total absorption cross section of CO$_2$ (see Figure~\ref{totabcsc}).
Using this cross section and S2K solar flux, the rate 
for O($^1$S) production  is 7.2 $\times$ 10$^{-7}$ s$^{-1}$  at 1 AU.

\subsubsection{CO}
The threshold wavelength for the dissociation of  CO molecule into neutral products in the ground state
 is 1117.8 \AA\  and in the metastable O($^1$D) and C($^1$D) is 863.4 \AA.
Among the O-bearing species discussed in this paper, CO has the highest dissociation energy of 11.1 
eV, while its ionization potential is 14 eV. 
  \cite{Huebner92} calculated cross sections for the photodissociative excitation 
 of CO producing O($^1$D) using branching ratios from  \cite{Mcelroy71} (cf. Fig.~\ref{phcsco1d-1}). 
Rates  for the production of O($^1$D) from CO molecule calculated by \cite{Huebner92} are 
 3.47 $\times$ 10$^{-8}$ and 
7.87 $\times$ 10$^{-8}$ s$^{-1}$ for solar minimum and maximum  conditions, respectively. 
 Using the cross section of \cite{Huebner92} and S2K model solar flux, our calculated 
rate for the O($^1$D) production from CO is 5.1 $\times$ 10$^{-8}$ s$^{-1}$ at 1 AU,  
which is 1.5 times higher than the solar minimum rate of \cite{Huebner92}. 
This difference in the calculated value is due to variation in the solar fluxes used in the two 
studies in wavelength region 600 to 800 \AA\ (cf.~Figure~\ref{solflx}).

We did not  find any  reports on the cross section 
for the production of O($^1$S) in photodissociation of the CO molecule.
According to \cite{Huebner79} the rate for this reaction can not be more than
 4 $\times$ 10$^{-8}$ s$^{-1}$. We have used this value in our model calculations.
This process can be an important source of O($^1$S) 
since the comet Hyakutake has a higher  CO abundance ($\sim$20\%).
 Using this photorate and CO abundance,  we will show that this reaction alone can contribute 
 up to a maximum of 30\% to the total  O($^1$S) production.

\subsection{Electron impact dissociation}
In our literature survey we could not find any  reported  cross section for the
production of O($^1$D) due to electron impact dissociation of H$_2$O. 
\cite{Jackman77} have assembled the experimental and theoretical 
cross sections for electron 
impact on important atmospheric gases in a workable analytical form. The cross sections for 
electron impact on atomic oxygen given by \cite{Jackman77} have been used to estimate  
emissions which leave the O atom in the metastable ($^1$D) state. The obtained ratios of 85\% in 
ground and 15\% in metastable state  are used for the atomic states of 
C and O produced in electron impact dissociation of H$_2$O, CO$_2$, and CO. It may be noted 
that the ground state to metastable state production ratio of 89:11 is observed 
for atomic carbon and atomic oxygen produced from photodissociation of CO \citep{Singh91}.
However, as shown later, the contributions of these 
electron impact processes to the total production of O($^1$D) are very small ($<$5\%).

 \cite{Kedzierski98} measured the cross section for electron impact dissociative 
excitation of H$_2$O producing O($^1$S), with overall uncertainty 
of 30\%. \cite{LeClair94} measured cross 
section for the production of O($^1$S) in dissociation of CO$_2$ by electron impact; they 
claimed an uncertainty of 12\% in their experimental cross section measurements. 
The cross section for fragmentation of
 CO into metastable O($^1$S) 
atom by electron impact is measured by \cite{LeClair94a}. 
These electron impact cross sections are also recommended by
 \cite{Mcconkey08}, and are used in our model for calculating the production
 rate  of O($^1$S) from H$_2$O, CO$_2$,  and CO.

\begin{figure}[h] 
\centering
  \noindent\includegraphics[width=20pc,angle=0]{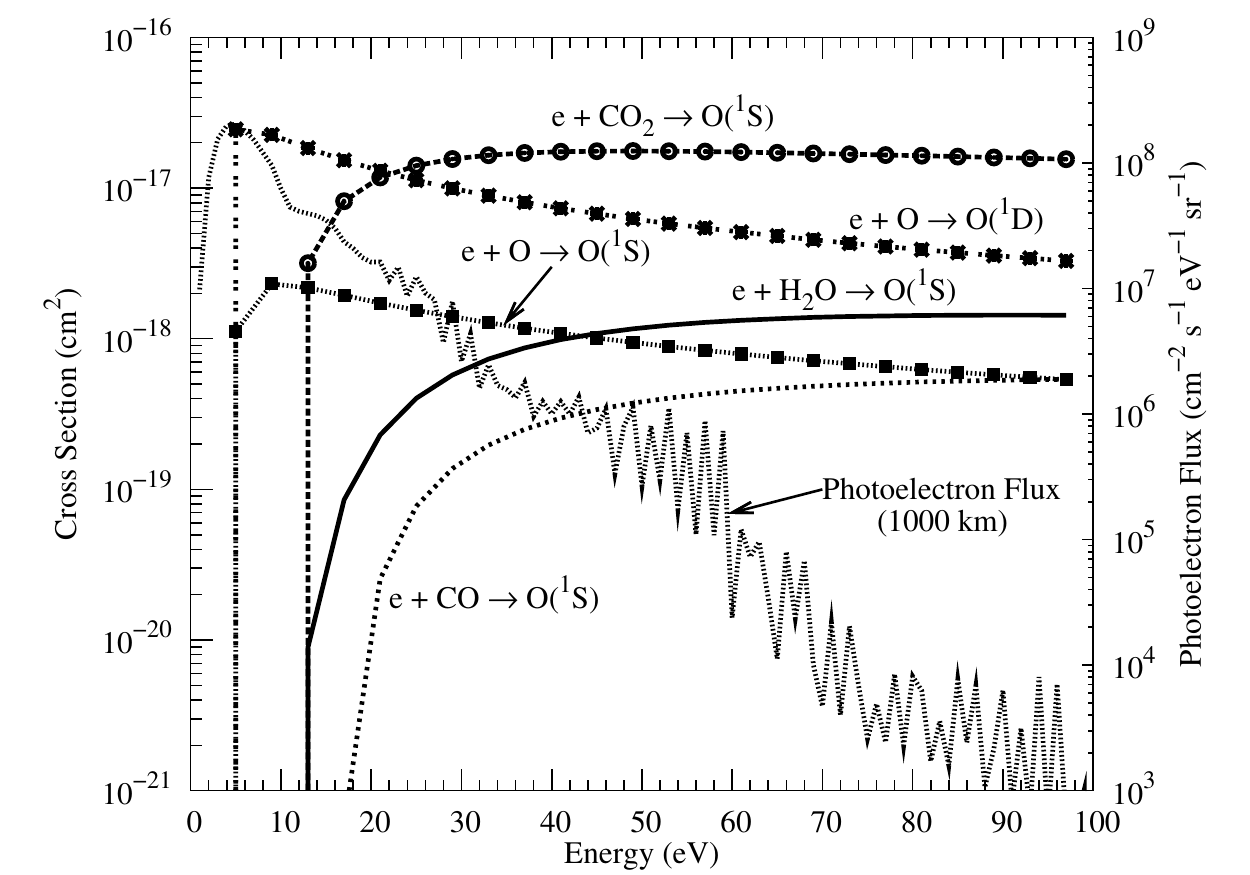}
\caption{Cross sections for the production of metastable O($^1$S) in electron impact dissociative 
excitation of H$_2$O, CO$_2$, and CO. Cross sections for electron impact excitation of O to O($^1$D) 
and O($^1$S) are also plotted.
Calculated steady state photoelectron flux at 1000 km cometocentric distance using S2K model solar 
flux on 1996 March 30 is also shown with scale on right side y-axis.}
\label{ecsco1s}
\end{figure}

Since the $^1$D and $^1$S are metastable states, the direct excitation of atomic oxygen
  by solar radiation is not an effective 
excitation mechanism. However the electron impact excitation of atomic 
oxygen can populate these excited metastable states, which is a major source of airglow
emissions  in 
the upper atmospheres of Venus, Earth, and Mars. 
We calculated the excitation rates for these  processes using 
 electron impact  cross sections from \cite{Jackman77}. 
 In calculating the photoelectron impact ionization rates of metastable oxygen states,
we calculated the cross sections  by changing the threshold energy parameter for ionization of neutral 
atomic oxygen in the analytical expression given by \cite{Jackman77}.
  The above mentioned  electron impact cross sections for the production of O($^1$S) from 
H$_2$O, CO$_2$, CO,  and O, used in the current model, are presented in 
Figure~\ref{ecsco1s} along with the calculated photoelectron flux energy spectrum
 at cometocentric distance of 1000 km.

\subsection{Dissociative recombination}
The total dissociative recombination rate for H$_2$O$^+$ reported by \cite{Rosen00} 
is 4.3 $\times$ 10$^{-7}$ cm$^{-3}$ s$^{-1}$ at 300 K.
 The channels of dissociative recombination 
 have also  been studied by this group.
  It was found that the dissociation process is dominated by three-body breakup (H + H + O) that occurs 
with a branching ratio of 0.71, while the fraction of two-body breakup (O + H$_2$)
  is 0.09, and the branching ratio for the formation of OH + H is 0.2.
The maximum kinetic energy of the  dissociative products forming atomic oxygen produced in ground 
state are 3.1 eV and 7.6 eV for the three-and two-body dissociation, respectively. Since the 
excitation energy required for the formation of metastable  O($^1$S) is 4.19 eV, the  three-body 
dissociation can not produce oxygen atoms in $^1$S state. However, the 
O($^1$D) atom can be produced in both,
the three-body and the two-body, breakup dissociation processes. 
To incorporate the contribution of H$_2$O$^+$ dissociative recombination in the production 
of O($^1$D) and O($^1$S), we assumed that 50\% of branching fraction 
of the total recombination in three-body and two-body breakups lead to the formation of O($^1$D)
 and O($^1$S)  atoms, respectively.
For dissociative recombination of CO$_2^+$, CO$^+$ and OH$^+$ ions  we assumed that the 
recombination rates are same for the production of both O($^1$D) and O($^1$S). 
We will show that these assumptions affects the calculated O($^1$S) and O($^1$D) densities
only  at larger ($\ge$ 10$^4$ km) cometocentric distances, but not in the inner coma.
Tables~\ref{tab-prlos1s} and \ref{tab-prlos1d} list the rates, along with the source reference, for these 
recombination reactions.

\section{Results and discussion}
\subsection{Production and loss of O($^1$S) atom}
\begin{figure}[h]  
\centering
  \noindent\includegraphics[width=20pc,angle=0]{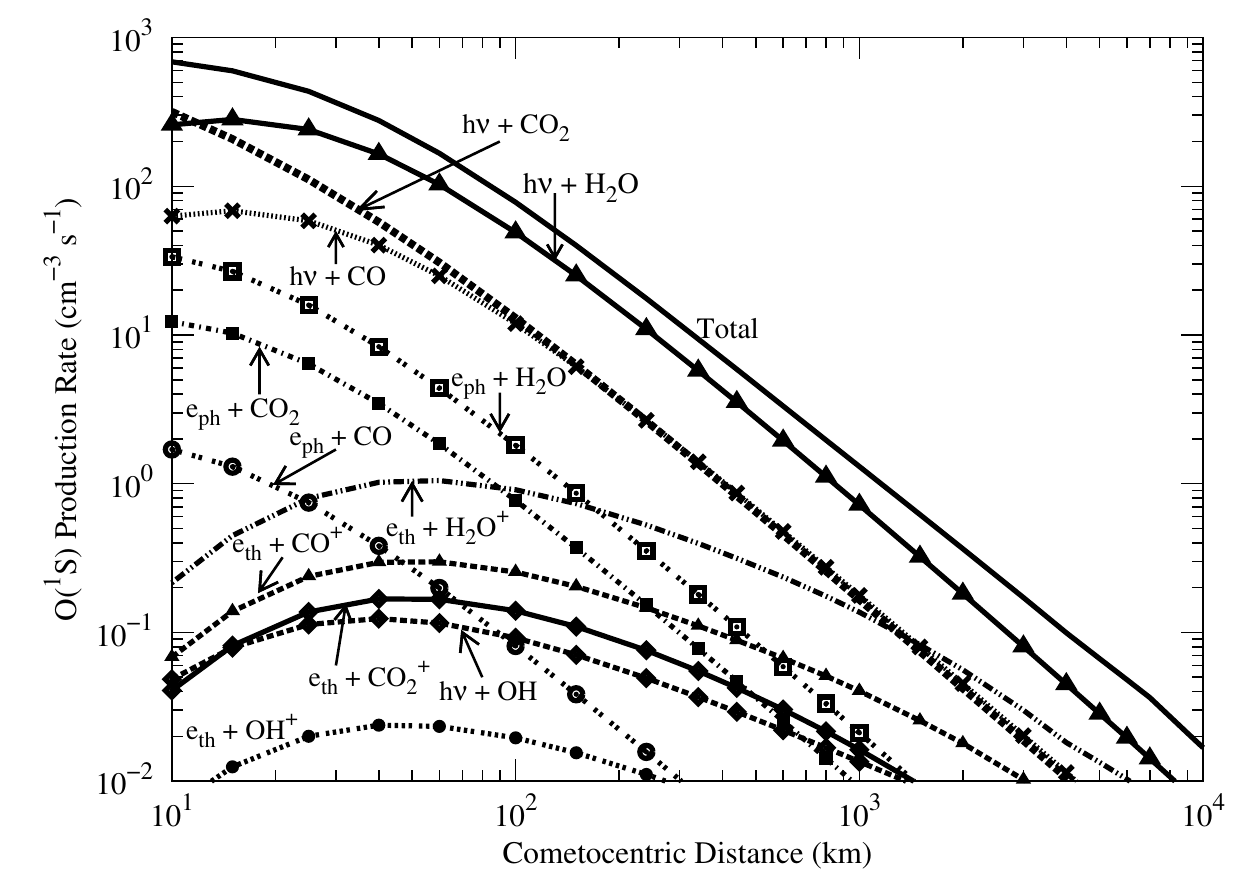}
\caption{Calculated radial profiles for major production mechanisms of O($^1$S) along with the 
total production profile. h$\nu$ = solar photon, e$_{ph}$ = photoelectron, and e$_{th}$ = thermal electron.}
\label{o1sprodr1}
\end{figure}

The calculated O($^1$S) production rate profiles for different processes 
in comet \hyak\ are presented in Figure~\ref{o1sprodr1}.
These calculations  are made under the assumption of  0.5\% yield of O($^1$S) from H$_2$O  at 1216 \AA\ 
solar H Lyman-$\alpha$ line and 1\% CO$_2$ relative abundance.  
 The major production source of O($^1$S) is the photodissociative excitation
 of H$_2$O throughout the cometary coma. However, very close to the nucleus, the photodissociative 
excitation of CO$_2$ is also an equally important process for the O($^1$S) production. 
Above 100 km, the photodissociative excitation of  CO$_2$ and CO makes an equal contribution in the  
production of O($^1$S). Since the cross section 
for electron impact dissociative excitation of H$_2$O, CO$_2$, and CO are small 
(see Figure~\ref{ecsco1s}), 
the contributions from electron impact dissociation to O($^1$S) production are smaller 
 by an order  of magnitude or more than that due to photodissociative excitation.  
At larger cometocentric distances ($>$2 $\times$ 10$^3$ km),
  the dissociative recombination of H$_2$O$^+$ ion is a significant 
 production mechanism for O($^1$S), whose contribution is
higher than those from photodissociative excitation of CO$_2$ and CO. 
The dissociative recombination of other ions do not make any significant contribution 
 to the production  of O($^1$S).

\begin{figure}[h]  
\centering
 \noindent\includegraphics[width=20pc,angle=0]{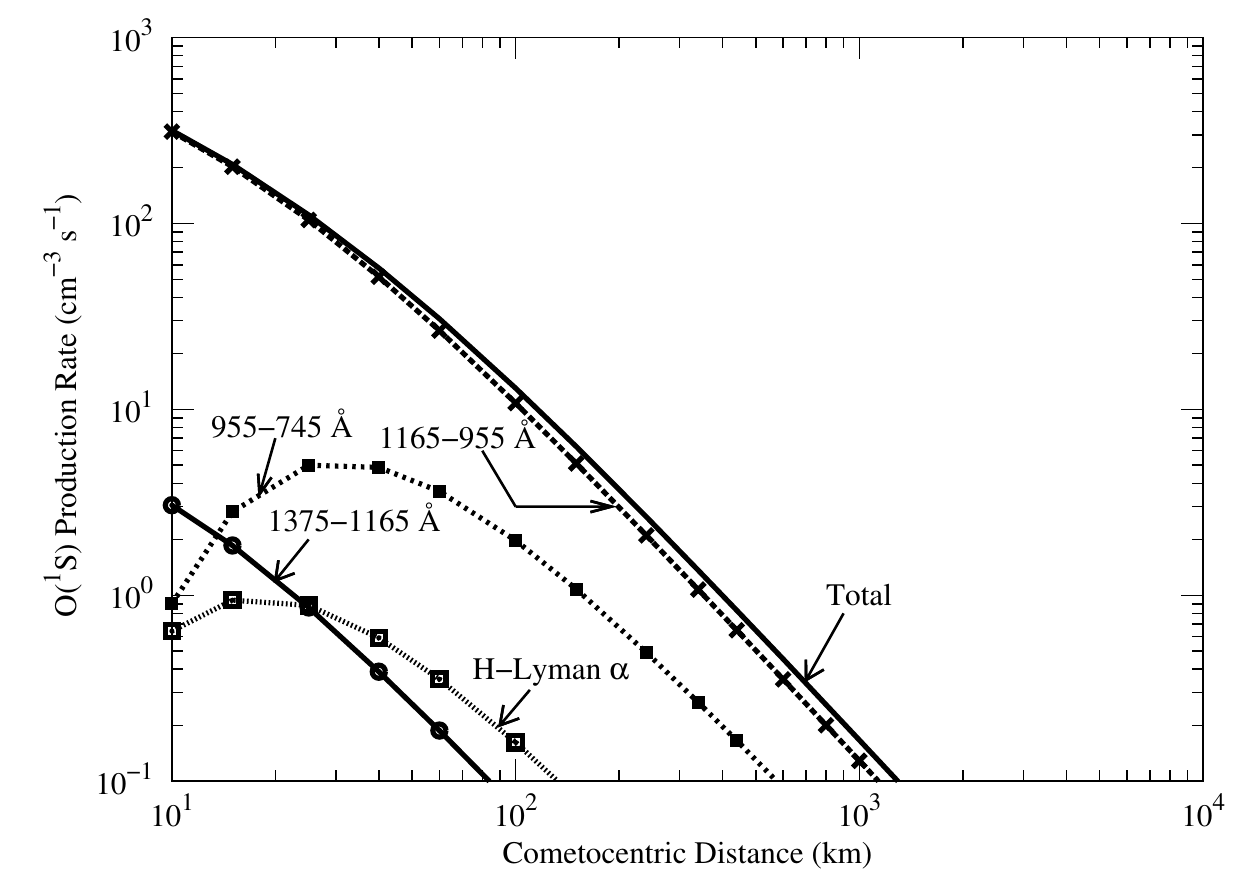}
\caption{Calculated  radial profiles for the  photodissociation of CO$_2$ producing 
 O($^1$S) at different wavelength bands.}
\label{o1s-pht-co2}
\end{figure}

\begin{figure}[h]  
\centering
  \noindent\includegraphics[width=20pc,angle=0]{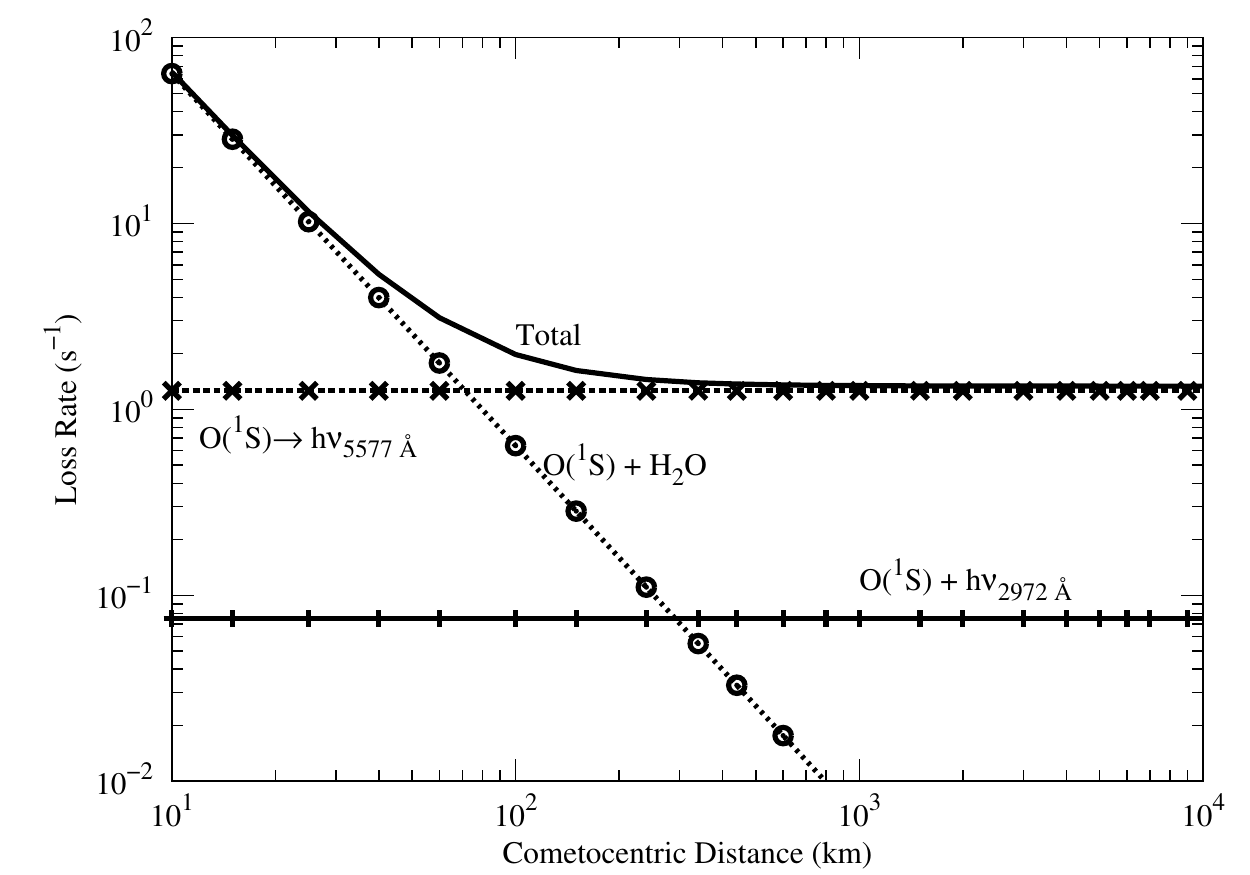}
\caption{Calculated radial profiles for the major loss mechanisms of O($^1$S) atom.}
\label{o1slos}
\end{figure}

In the inner coma, the  calculated production rates of O($^1$S) 
via photodissociative excitation is CO$_2$ at various wavelengths are presented in 
Figure~\ref{o1s-pht-co2}. The major 
production of O($^1$S)  occurs in the wavelength 
region 955--1165 \AA\
where the average cross section  is $\sim$2 $\times$ 10$^{-17}$ cm$^{-2}$ 
(cf. Fig.~\ref{phcsco1d-1})  and the average solar flux  is $\sim$1 $\times$ 
10$^{9}$ photons cm$^{-2}$ s$^{-1}$ (cf. Fig.~\ref{solflx}).
The calculated loss rate profiles of O($^1$S) for major processes
 are presented in Figure~\ref{o1slos}.
 Close to the nucleus ($<$50 km), quenching  by H$_2$O is the main loss mechanism 
for  metastable O($^1$S). Above 100 km, the
radiative decay of O($^1$S) becomes the dominant loss process. The contributions from  
other loss processes are  orders of magnitude smaller and hence are not shown 
in Figure~\ref{o1slos}.

\subsection{Production and loss of O($^1$D) atom}
\begin{figure}  
\centering
  \noindent\includegraphics[width=20pc,angle=0]{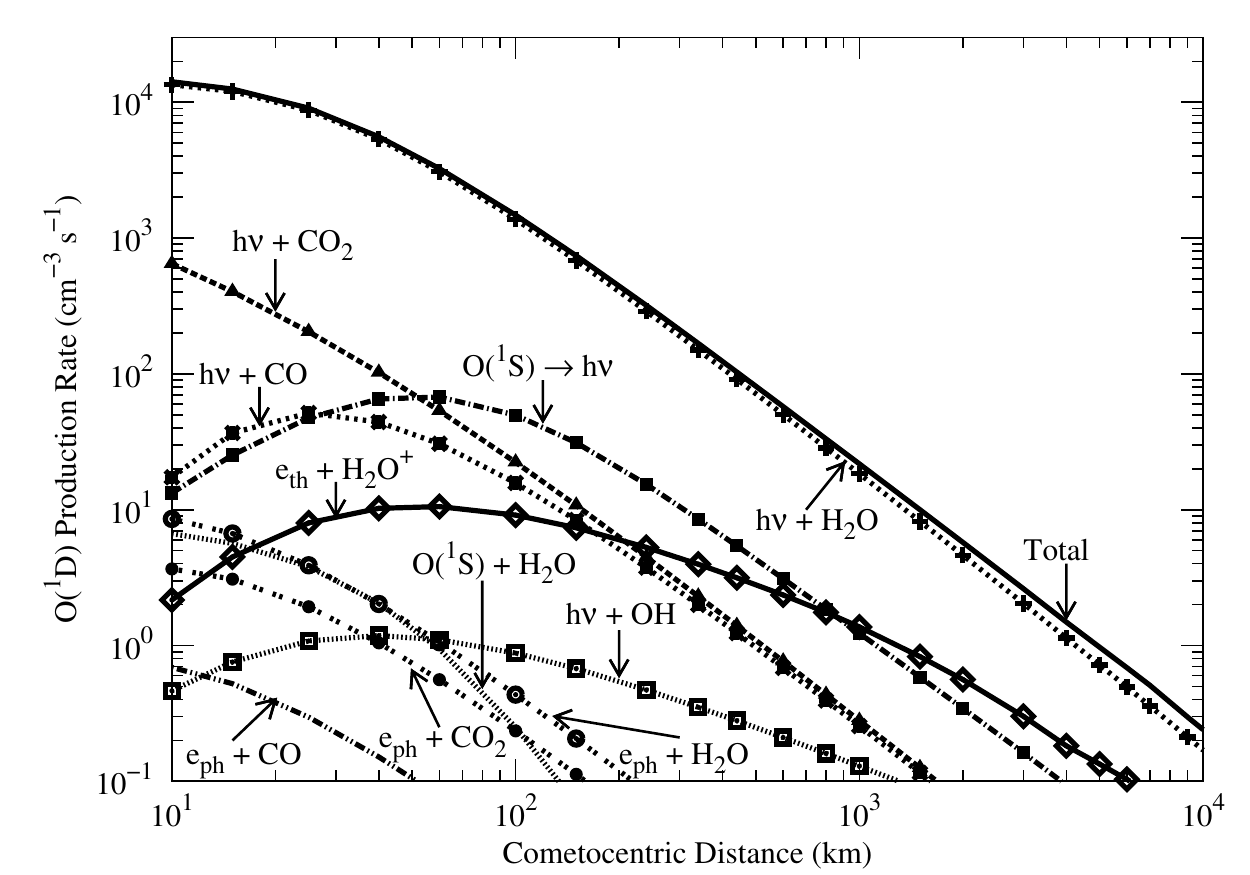}
\caption{Calculated  radial profiles for the major production mechanisms of O($^1$D)
 along with total O($^1$D)
production rate profile. h$\nu$ = solar photon, e$_{ph}$ = photoelectron, and 
e$_{th}$ = thermal electron.}
\label{o1dprodr1}
\end{figure}

\begin{figure} 
\centering
   \noindent\includegraphics[width=20pc,angle=0]{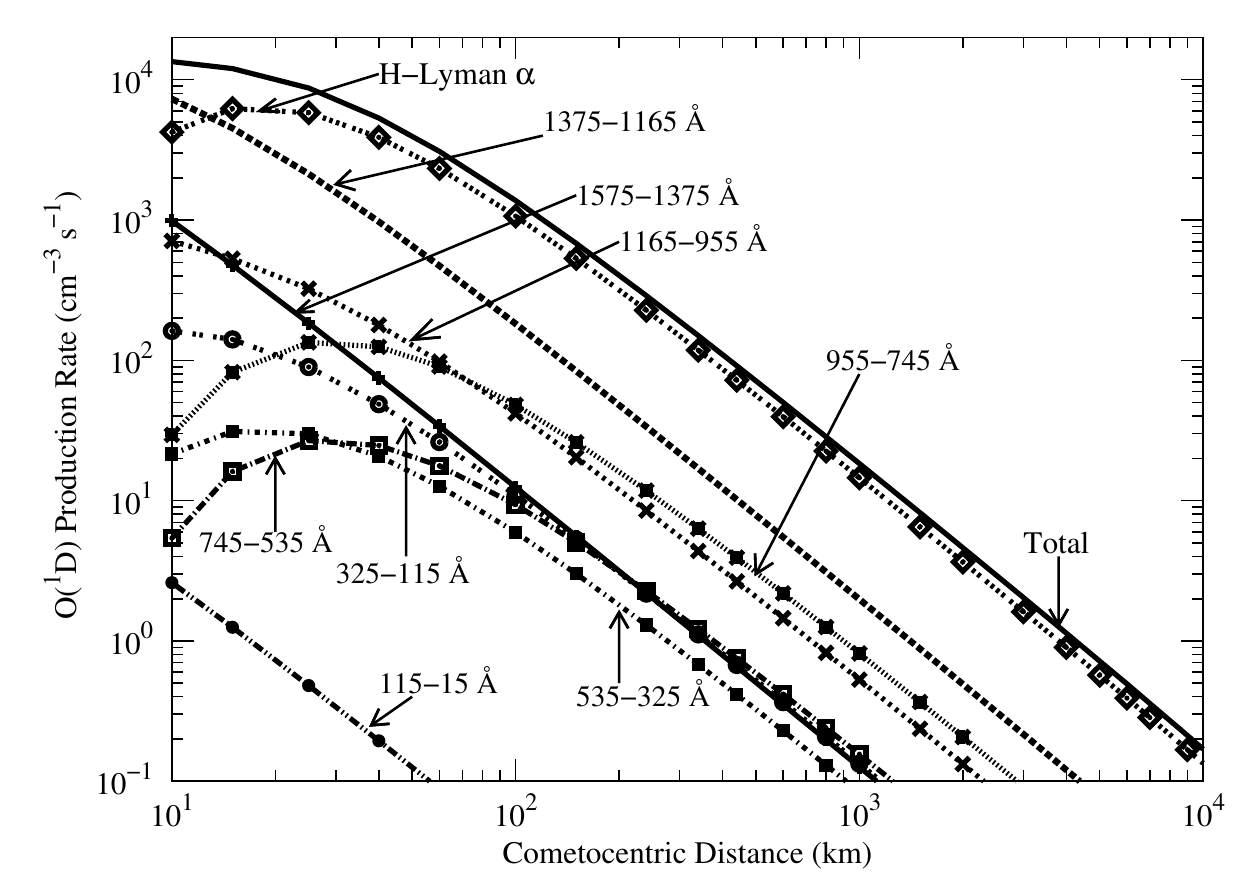}
\caption{Calculated  radial profiles for the  photodissociation of H$_2$O producing 
 O($^1$D) at different wavelength bands.}
\label{o1d-pht-h2o}
\end{figure}

The production rates as a function of cometocentric distance for various excitation mechanisms 
of the O($^1$D) are shown in Figure~\ref{o1dprodr1}. The major source of O($^1$D) production in the 
inner coma  is photodissociation of H$_2$O.
The wavelength dependent production rates of O($^1$D) from H$_2$O are presented in 
Figure~\ref{o1d-pht-h2o}.  The O($^1$D) production in photodissociation of  
H$_2$O  is governed by solar radiation at H Lyman-$\alpha$ (1216 \AA) wavelength.
However, very close to the nucleus, the production of O($^1$D) is 
 largely due to photons in the wavelength region 1165--1375 \AA.  Since the average absorption 
cross section of H$_2$O  decreases in this wavelength region by an order of magnitude,  
the optical depth 
at  wavelengths greater than 1165 \AA\ is quite small (see Figure~\ref{totabcsc}). 
Hence, these photons are able to  travel deeper into the coma unattenuated, thereby 
reaching close to the nucleus where they dissociate 
H$_2$O  producing O($^1$D). Thus, at the surface of cometary nucleus the production 
of O($^1$D) is controlled by the solar radiation in this wavelength band. In high production rate comets, 
the  production   of  O($^1$D) near nucleus  would be governed by 
solar photons in this wavelength region. The production of O($^1$D) from H$_2$O by 
solar photons from other wavelength regions is smaller by more than an order of magnitude.

\begin{figure}  
\centering
  \noindent\includegraphics[width=20pc,angle=0]{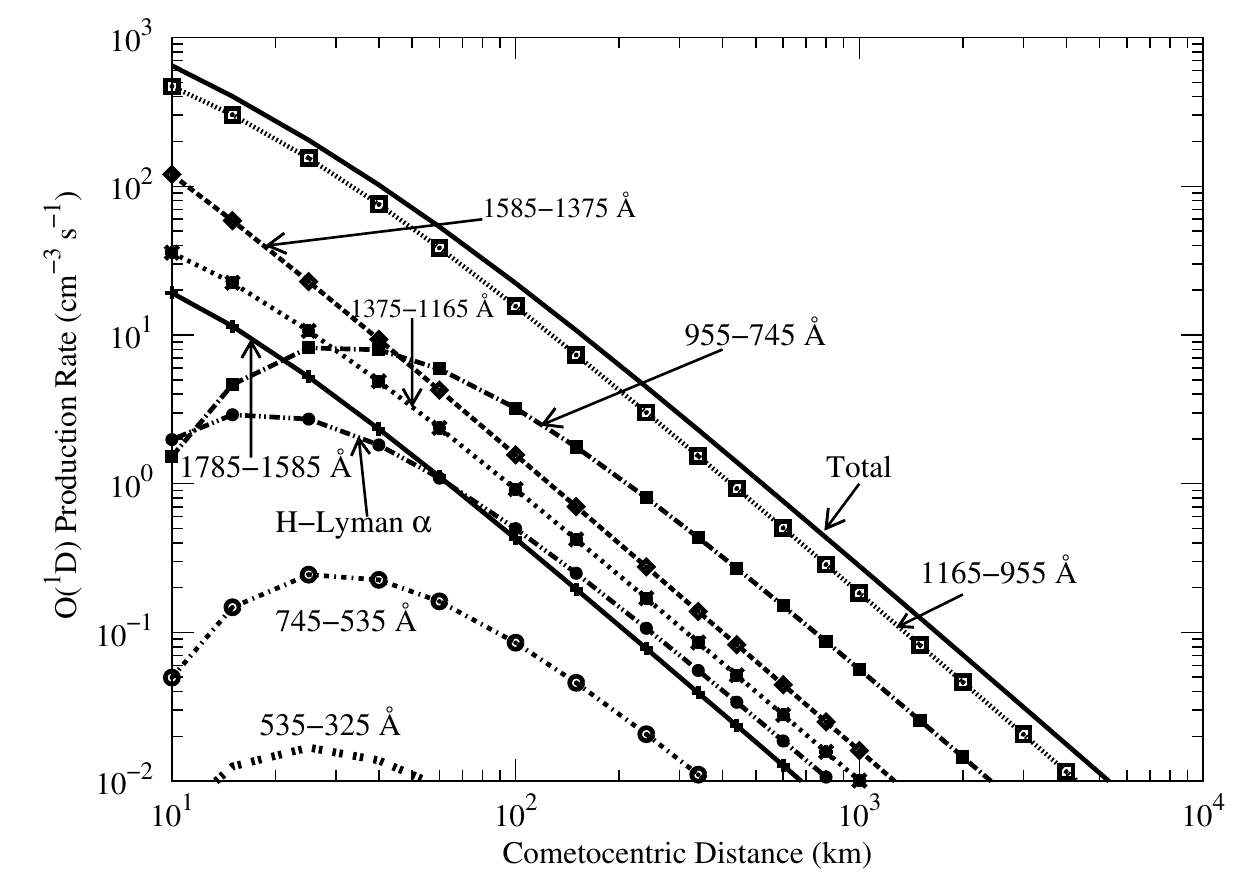}
\caption{Calculated  radial profiles for the  photodissociation of CO$_2$ producing 
 O($^1$D) at different wavelength bands.}
\label{o1d-pht-co2}
\end{figure}

After photodissociative excitation of H$_2$O, the next significant O($^1$D)  production process
 at radial distances below 50 km  is the photodissociative excitation of CO$_2$.
Above 50 km to about 1000 km, the radiative decay of O($^1$S), and at radial distances above 1000 km
the dissociative recombination of H$_2$O$^+$, are the next potential sources of the O($^1$D) 
 (see Figure~\ref{o1dprodr1}). 
The calculated wavelength dependent production rates of O($^1$D) for 
photodissociation of CO$_2$  are shown in Figure \ref{o1d-pht-co2}. Solar 
 radiation in the wavelength region 1165--955 \AA\ dominates the  
 O($^1$D) production. Since the cross section for the production of O($^1$D) 
due to photodissociation of  CO$_2$ is more than an order of magnitude
higher in this wavelength region compared to cross section  at
other wavelengths (see  Figure~\ref{phcsco1d-1}), the solar radiation in 
this wavelength band mainly controls the formation of O($^1$D) from CO$_2$. 
Other potential contributions are made by solar photons in the wavelength  band 1585--1375 \AA\  
at distances $<$50 km, and 955--745 \AA\ at radial distances $>$100 km.
 Since the CO$_2$ absorption cross section  around 
1216 \AA\ is smaller by more than two orders of magnitude compared to its maximum value, 
the solar radiation at H Ly-$\alpha$ is not an efficient source of 
 O($^1$D) atoms.

 \cite{Zipf69}  measured the total rate coefficient for the quenching of O($^1$S) by H$_2$O  as 
3 $\times$ 10$^{-10}$ cm$^{3}$ s$^{-1}$. The primary channel in  quenching mechanism is   
the production of two OH atoms.
The production of O($^1$D) is  also a possible channel whose rate coefficient
 is not reported in the literature.
 Hence, we assumed that 1\% of total  rate coefficient can lead to the formation of O($^1$D) 
in this quenching mechanism. However, this assumption has no implications on 
the O($^1$D) production 
since the total contribution due to O($^1$S) is about three orders of magnitude smaller than 
the major production process of O($^1$D).

\begin{figure}  
\centering
  \noindent\includegraphics[width=20pc,angle=0]{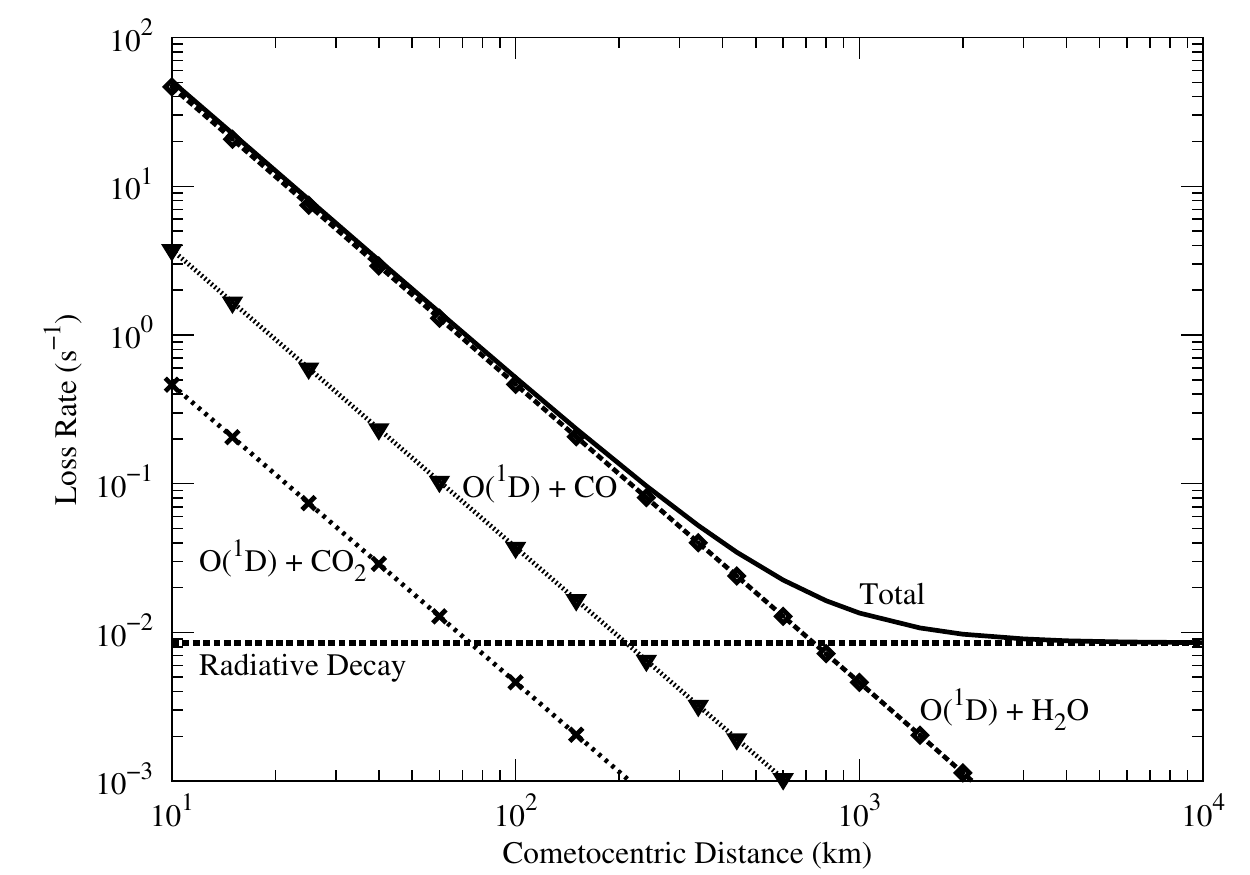}
\caption{Calculated radial profiles for major loss mechanisms of O($^1$D) atom.}
\label{o1dlos}
\end{figure}

The calculated loss rate profiles of O($^1$D) are presented in Figure~\ref{o1dlos}.
Below 1000 km, the O($^1$D) can be quenched by various cometary species. The quenching by H$_2$O  
is the major loss mechanism for O($^1$D) below 500 km. Above 2 $\times$ 10$^3$ km 
 radiative decay is the dominant loss process for O($^1$D).

\subsection{Calculation of green and red-doublet emission intensity}
\begin{figure}  
\centering
  \noindent\includegraphics[width=20pc,angle=0]{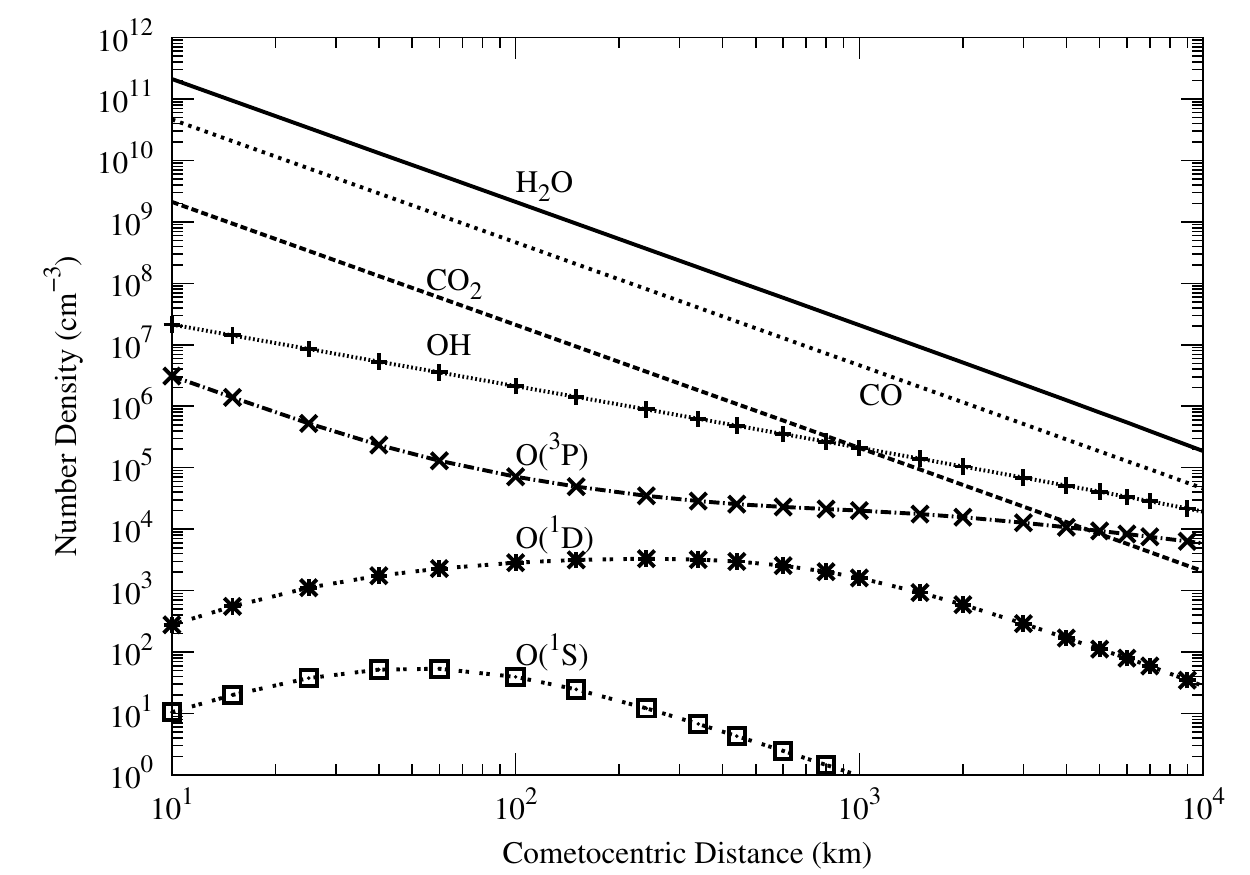}
\caption{Calculated number density  profiles of O($^1$S), O($^1$D), O($^3$P), and OH, 
along with those of H$_2$O, CO, and CO$_2$.}
\label{nubden}
\end{figure}

Using the calculated production and loss rates due to various processes 
mentioned above, 
and assuming photochemical equilibrium,  we computed the number density 
of O($^1$S) and O($^1$D) metastable atoms. The calculated number densities are presented
in Figure~\ref{nubden}.
The O($^1$D) density profile shows a broad peak around 200--600 km.
 But, in the case of O($^1$S), the density peaks at much lower radial distances of $\sim$60 km.
The number densities of O($^1$D) and O($^1$S) are converted into emission rate profiles 
 for the red-doublet and green line emissions, respectively, by  multiplying with Einstein transition 
probabilities as
\begin{eqnarray}
V_{(6300+ 6364)}(r) =A_{(6300+ 6364)} \times [O^1D(r)] \nonumber \\ = A_{(6300+ 6364)} 
\frac{\sum_{i=1}^k P_i(r)}{\sum_{i=1}^k L_i(r) +  A(^1D) }
\end{eqnarray}

and
\begin{eqnarray}
V_{(5577)}(r) = A_{(5577)} \times [O^1S(r)] \nonumber \\ = A_{(5577)} 
\frac{\sum_{i=1}^k P_i(r)}{\sum_{i=1}^k L_i(r) +  A(^1S) }
\end{eqnarray}
Where $[O^1S(r)]$ and $[O^1D(r)]$ are the calculated number density  for the corresponding 
 production rates $P_i(r)$ and loss frequencies $L_i(r)$  for  O($^1$S) and O($^1$D), respectively.
$A(^1D)$ and $A(^1S)$ are the total Einstein spontaneous emission coefficients for red-doublet 
and green line emissions. 
 Using the emission rate profiles, the line of sight intensity of green and red-doublet emissions
 along the projected distance $z$ is calculated as 
\begin{equation}
 I(z) = 2  \int_{z}^{R}V_{(5577,\ 6300+6364)} (s)ds
\end{equation}
where $s$ is the abscissa along the line of sight, and V$_{(5577,\ 6300+6364)}(s)$ is 
 the emission rate for the  green or red-doublet emission.
 The maximum limit of integration $R$ is taken as 10$^5$ km.
The calculated  brightness profiles of 5577 and  6300 \AA\ emissions are presented in
 Figure~\ref{o1so1d-cmp}.
These brightness profiles are then averaged over the projected area 
  corresponding to the  slit dimension 1.2$''$ $\times$ 8.2$''$  centred 
on the nucleus of comet \hyak\ for the observation on 30 March 1986 \citep{Cochran08}.  
The G/R ratio  averaged over the slit  is also calculated.

\subsection{Model results}
\cite{Morrison97} observed the green and red-doublet emissions on comet \hyak\
in the high resolution optical spectra obtained  on 1996 March 23 and 27
 and found the G/R ratio in the range 0.12--0.16.
\cite{Cochran08}  observed the 5577 and 6300~\AA\ line emissions on this 
 comet  on 1996 March 9 and 30, with the G/R ratio as 0.09 for 9 March observation.
We  calculated the G/R ratio  by varying the yield  
for  O($^1$S) production in photodissociation of H$_2$O at 1216 \AA\ (henceforth refer 
to as O($^1$S) yield).
 Since CO$_2$ is not observed in this comet, we assumed that a 
minimum  1\%  CO$_2$ is present  in 
the  coma. However, we also carried out calculations for  0\%, 
3\% and 5\% CO$_2$ abundances in the comet. 
We calculated the contributions of different production processes in the formation of 
O($^1$S) and O($^1$D) at three different projected distances of 10$^2$, 10$^3$, and 10$^4$ km 
from the nucleus for the above mentioned CO$_2$ abundances and the O($^1$S) yield values
varying from 0\% to 1\%. These calculations are   presented in Table~\ref{tabprj-yld}.
The percentage contribution of major production processes in the projected  field of view 
for the green and red-doublet emissions are also calculated. The G/R 
ratio is calculated after averaging the intensity  over the projected area 
165 $\times$ 1129 km which corresponds to the dimension of slit used in the observation made 
by \cite{Cochran08} on 1996 March 30.
These calculated values are presented in Table~\ref{tabprj-slit}. 

Taking  1\% CO$_2$ abundance and 0\% O($^1$S) yield, the calculated percentage 
contributions of major production processes of O($^1$S) and  O($^1$D) atoms 
are presented in Table~\ref{tabprj-yld}.
Around 60 to 90\% of the  O($^1$D) is produced from photodissociation of H$_2$O. Contributions 
of photodissociative excitation of CO$_2$ and CO in the production of O($^1$S) and O($^1$D) are
15 to 40\% and 1\%, respectively.
Around 10$^4$ km projected distance, the photodissociative excitation 
of OH ($\sim$20\%) and the dissociative recombination of H$_2$O$^+$ ($\sim$30\%) are also significant
production processes for O($^1$S) atoms. But, the contributions from these processes in 
  O($^1$D) production is around 10\% only.

For CO$_2$ abundance of 1\% and O($^1$S) yield of 0.2\%, the calculations presented 
in Table~\ref{tabprj-yld} show that the photodissociation of H$_2$O  contribute  around 
20 to 40\% in the production of  O($^1$S) and 60 to 90\% in the production of O($^1$D)  
atom. The next major source of O($^1$S) production is the
photodissociation of CO$_2$ and CO with each
contributing $\sim$10 to 25\%. 
 The relative contributions from photodissociation of parent species H$_2$O, CO$_2$, and CO 
to O($^1$S) and O($^1$D) production decreases with increase in projected distance from the 
nucleus. At 10$^4$ km projected distance, the
photodissociation of OH contribute 15\% and 8\% to the production of O($^1$S) and O($^1$D) atoms,
 respectively.
Above 1000 km projected distance, the contribution of H$_2$O$^+$
dissociative recombination to  O($^1$S) production is around 20\%.
 The production of O($^1$D)  atom is mainly via photodissociation of H$_2$O, but 
around 10$^4$ km  the dissociative recombination of H$_2$O$^+$ ion is 
also a significant production process contributing around 12\%. At 10$^4$ km,  
 dissociative recombination of OH$^+$ also contribute around 10\% to the total O($^1$D) production, 
 which is not shown in  Table~\ref{tabprj-yld}, and this value is independent of O($^1$S) yield 
 or  CO$_2$ abundance. Radiative decay of O($^1$S) is a minor ($\le$5\%) production process 
in the formation of  O($^1$D). 

 We also calculated the relative contributions of different processes in the 
formation of green and red line emissions in the slit projected field of view, which are presented 
in Table~\ref{tabprj-slit}. For the above case, 
the photodissociation of H$_2$O contribute around 35\%, while the photodissociation of CO$_2$ and CO 
 contribute  23\% and 22\%, respectively, to the production of green line emission. 
The contribution of dissociative recombination of H$_2$O$^+$ ions is around 10\%.
 The major production process of red lines is photodissociation of H$_2$O (90\%); 
the  dissociative recombination of H$_2$O$^+$ and radiative decay of O($^1$S) atom
are minor ($\le$5\%) production processes. With the O($^1$S) yield of 0.2\% and 1\% CO$_2$ abundance, 
the slit-averaged G/R ratio is found to be 0.11.

 When  the  O($^1$S) yield is increased  
to 0.5\% with 1\% CO$_2$ abundance (see Table~\ref{tabprj-yld}), the contribution from 
photodissociative excitation of H$_2$O to the O($^1$S) production  is increased, with value 
varying from 35 to 60\%, while the contribution to O($^1$D) 
 production is not changed. In this case, the contribution from photodissociation 
of CO$_2$ and CO to the O($^1$S) production is reduced (values between 10 to 15\%). 
The contributions from other processes are not changed significantly. 
Table~\ref{tabprj-slit} shows that in this case 
 around 60\% of green line  in the slit projected field 
of view is produced  via photodissociation of H$_2$O, while the contributions from 
photodissociation of CO$_2$ and CO are around 15\% each. The main (90\%) production of 
red-doublet emission is through photodissociation of H$_2$O. The slit-averaged G/R
 ratio is 0.17. 

On further increasing the O($^1$S) yield  to 1\% with CO$_2$ abundances of 1\%, the contribution of 
photodissociation of H$_2$O to O($^1$S) atom production is further increased 
(values between 50 to 75\%)  while the 
contribution from photodissociation of CO$_2$ and CO is decreased to around 
10\% each (cf. Table~\ref{tabprj-yld}). The contributions from
other processes are not affected compared to the previous case.  
As seen from Table~\ref{tabprj-slit}, in this case the contribution of photodissociation of
 H$_2$O  to green line is around 75\% in the 
slit projected field of view,  while contributions from photodissociation of CO$_2$ and CO 
are decreased to 10\% each. The calculated  G/R ratio is 0.27 (Table~\ref{tabprj-slit}).

We also evaluated the effect of CO$_2$  on the red-doublet and green line emissions
by varying its abundance to 0\%, 3\% and 5\%. The calculated percentage contribution of major processes
 along the projected distances and in the slit projected field of view
 are presented in Tables~\ref{tabprj-yld} and \ref{tabprj-slit}, respectively.
 In the absence of CO$_2$, the contributions 
from H$_2$O, H$_2$O$^+$  and CO in O($^1$S) production are increased by 
 $\sim$10\% (cf. Tables~\ref{tabprj-yld} and \ref{tabprj-slit}).
 Taking 0\% O($^1$S)  yield and by increasing  CO$_2$ relative abundance from 1 to 3\%, 
 the percentage contributions for O($^1$S) from  photodissociative excitation of CO$_2$ (CO)
is increased (decreased) by 50\%.  The contribution from  H$_2$O to 
O($^1$D) production is not changed.

The calculations presented in Tables~\ref{tabprj-yld} and \ref{tabprj-slit} depict that
 the contributions of various processes are significant
in the production  O($^1$S)  atom, whereas  photodissociative 
excitation of H$_2$O is the main production  process for  O($^1$D) atom.
Since  comet \hyak\ is rich in CO (abundance $\sim$22\%) compared to other comets, the  
contribution from CO photodissociation to O($^1$S) production  
is significant (10--25\%). 
In the case of a comet having CO abundance less than 20\%, the major production source of 
metastable  O($^1$S) atom would be  photodissociation of H$_2$O and CO$_2$.

\subsection{Comparison with observations}
In 1996 March, the green and red-doublet emissions were observed 
in comet \hyak\  from two ground-based observatories 
\citep{Morrison97,Cochran08}. Each observatory determined 
the G/R ratio using different slit size. Using a circular slit, the projected radial distance over 
the comet for  \cite{Morrison97} observation on 
 March 23 and March 27 varied from 640 to 653 km, while for \cite{Cochran08} observation, using a 
rectangular slit,  the 
projected area was 480 $\times$ 3720 km on March 9 and 165 $\times$ 1129 km on March 30. 
The clear detection of both green and red-doublet emissions and determination of the
 G/R ratio could be 
done for March  9 and March 23 observations only \citep{Cochran08,Morrison97}. The observed G/R
 ratio was 0.09 and 0.12 to 0.16 for the observation on March 9 and March 23, respectively.

\begin{figure}  
 \centering
  \noindent\includegraphics[width=20pc,angle=0]{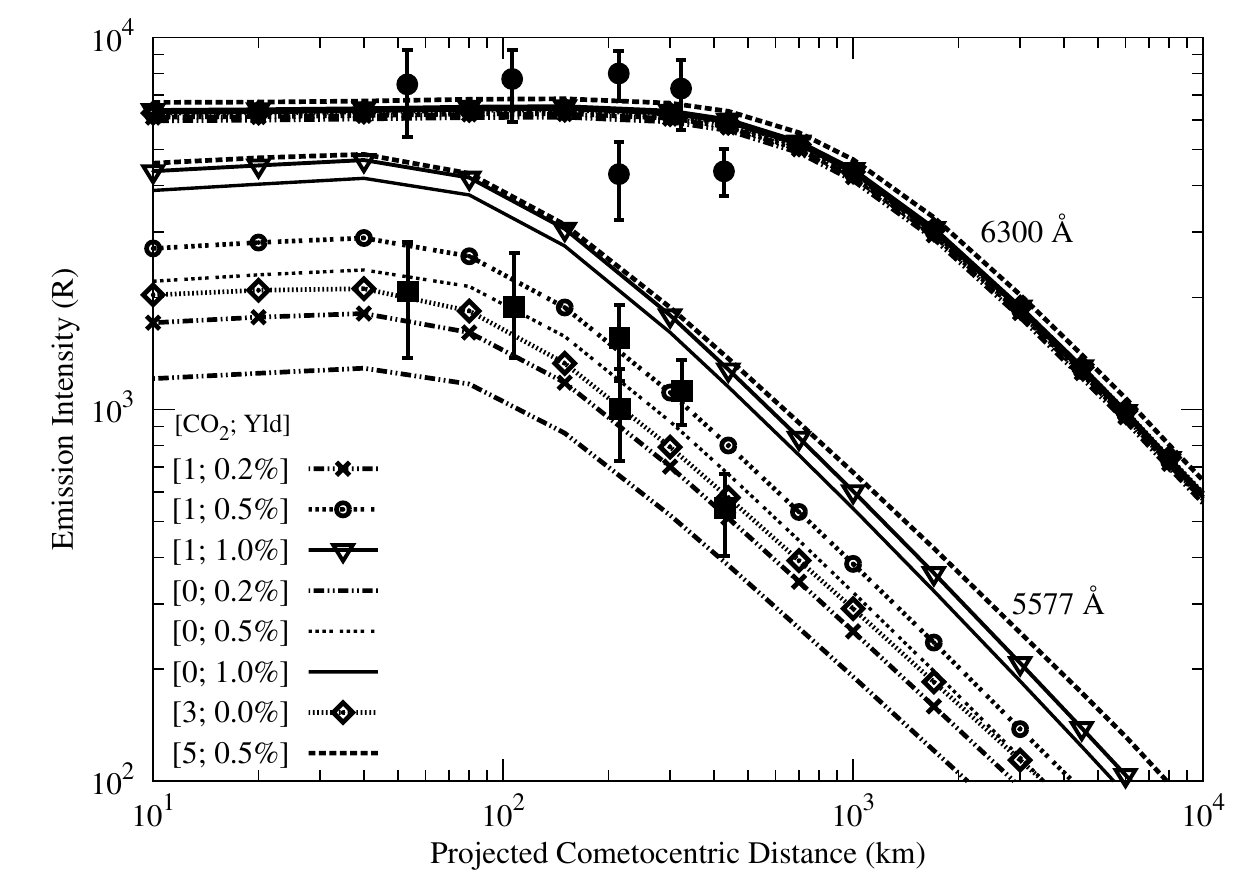}
\caption{Calculated emission brightness profiles along projected distances for 5577 \AA\
(green)
and 6300 \AA\ (red) line emissions  for different  CO$_2$ relative abundance [CO$_2$] and 
 yield [Yld] for O($^1$S) production in photodissociation of H$_2$O. 
The green and red emission intensities at 
different projected distances observed  on  March 30  taken  from \cite{Cochran08} 
 are also shown (filled symbols with error bars) for comparison with the calculated values.}
\label{o1so1d-cmp}
\end{figure}

\begin{figure}  
 \centering
  \noindent\includegraphics[width=20pc,angle=0]{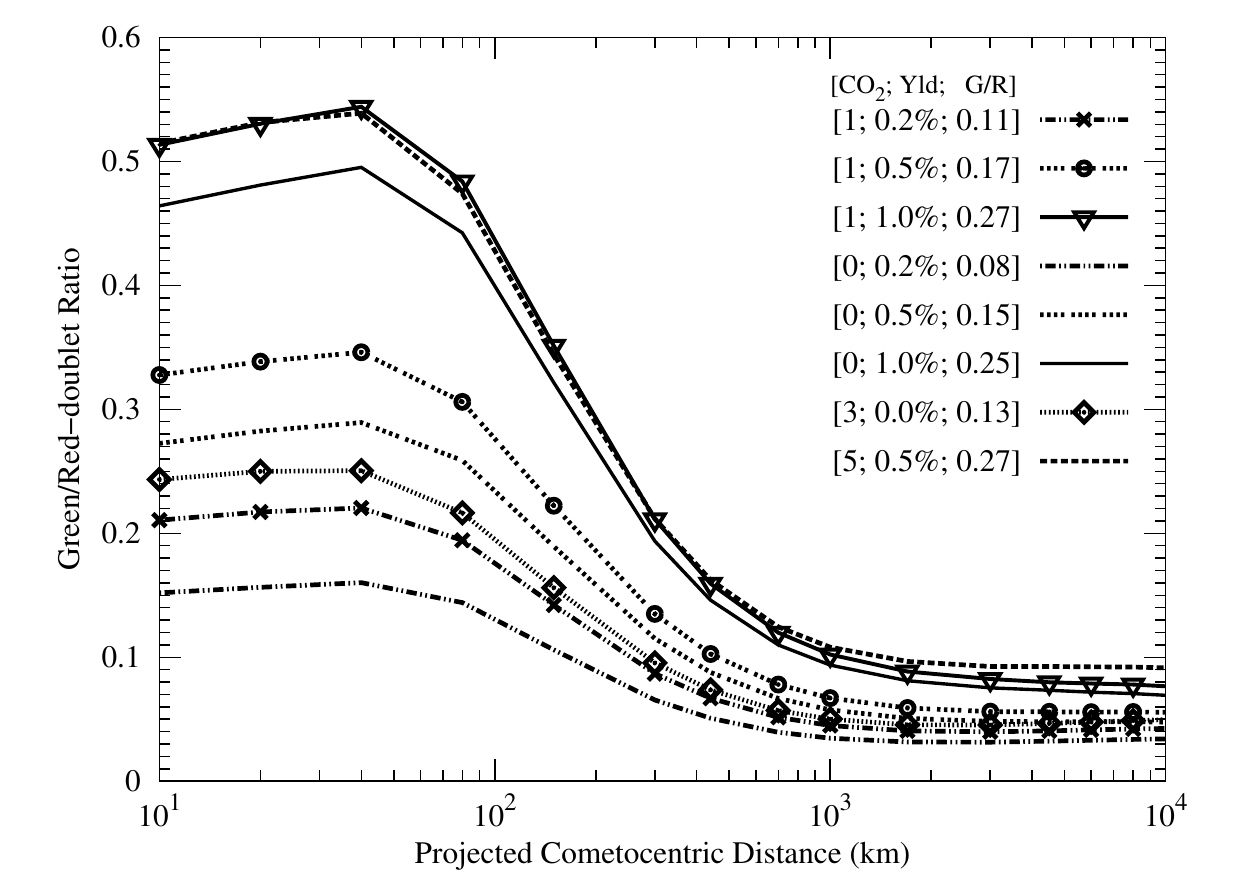}
\caption{Calculated green to red-doublet intensity ratio along  
projected distances for different  CO$_2$ relative abundance [CO$_2$] and 
 yield [Yld] for O($^1$S) production in photodissociation of H$_2$O. 
 G/R = calculated green to red-doublet intensity 
ratio averaged over slit projected size 165 $\times$ 1130 km for \hyak\ on 1996 March 30.}
\label{ratio-cmp}
\end{figure}

Making a very high resolution (R = 200,000) observation of comet \hyak\ on 1996 March 30, 
\cite{Cochran08} obtained radial profiles of 5577  
and 6300 \AA\ lines. In Figure~\ref{o1so1d-cmp}
 we have compared the model  calculated intensity profiles of 6300 and 5577 \AA\ lines 
at different projected 
cometocentric distances with the observation of \cite{Cochran08}. The calculated G/R ratio
along projected distance is  shown in Figure~\ref{ratio-cmp}. The 6300 \AA\ emission
shows a flat profile upto $\sim$500 km, whereas the 5577 \AA\ green 
line starts falling off beyond 100
km. This is because of the quenching of O($^1$S) and O($^1$D) by H$_2$O in the inner most coma 
(cf. Figures \ref{o1slos} and \ref{o1dlos}), thereby making both the production and loss mechanisms 
being controlled by H$_2$O. Above these distances, the emissions are mainly controlled by the radiative 
decay of  $^1$S  and  $^1$D  states of oxygen atoms.

Similar to the calculations presented in Tables~\ref{tabprj-yld} and 
\ref{tabprj-slit}, in  Figures~\ref{o1so1d-cmp} and~\ref{ratio-cmp} we present the red 
and green line intensity profiles
and the G/R ratios, respectively, for different contributions of O($^1$S)
 yield and CO$_2$ abundances. 
Since photodissociative excitation of H$_2$O is the main production process for
  O($^1$D) atom, the red line intensity is almost independent of the variation 
in O($^1$S) yield and CO$_2$ abundance.
In the case of 0\% CO$_2$ abundance, the best  fit to the observed green line profile is 
obtained when the O($^1$S) yield is $\sim$0.5\% ($\pm$ 0.1\%), where the G/R ratio
varied from 0.06 to 0.26 (cf. Figure~\ref{ratio-cmp}) and the slit-averaged 
G/R ratio for  March 30
observation is 0.15 (cf. Table~\ref{tabprj-slit}).
 The shape of green line profile cannot be explained with 1\% or 0\% O($^1$S)
yield, while the case for 0.2\% O($^1$S) yield can be considered as somewhat consistent with 
the observation. For this case, the value of G/R ratio shown in Figure~\ref{ratio-cmp}
 is found to vary over a large range of 0.54 to 0.02.

When we consider 1\% CO$_2$ in the comet, the best-fit green profile is obtained when the O($^1$S) yield 
is $\sim$0.2\%. The case for 0.5\% O($^1$S)  yield also provides the green line profile
 consistent with the observation. In both these cases the G/R ratio varies between 0.32 and 
0.04 over the cometocentric projected distances of 10 to 10$^4$ km. 
The calculated 5577 \AA\ profiles for O($^1$S) yield of 0\% 
and 1\% are inconsistent with the observed profile.

 In Figure~\ref{o1so1d-cmp} we also show a calculated
profile for a case when the CO$_2$ abundance is 3\% while the O($^1$S) yield is 0\%  (i.e., no O($^1$S) is 
 produced in photodissociation of H$_2$O). The calculated 5577 \AA\ green line profile 
shows a good fit to the observed profile: suggesting that even a small abundance of CO$_2$ 
is enough to produce the required O($^1$S). This is because the CO$_2$ is about an order of magnitude
more efficient in producing O($^1$S) atom than H$_2$O in the photodissociation process 
(see Table~\ref{tab-prlos1s}). However, since O($^1$S) would definitely be produced in the 
photodissociation
of H$_2$O, and that the CO$_2$ would surely be present in comet (though  in smaller abundance),
 the most consistent value for the O($^1$S) yield would be around 0.5\%. Assuming 
5\% CO$_2$ and 0.5\% O($^1$S) yield, the calculated green line emission profile  
 is inconsistent with the observation (cf. Figure~\ref{ratio-cmp}). 
In this case, the calculated G/R ratio shown in Figure~\ref{ratio-cmp} is found to vary between 
0.24 and 0.05.

From the above calculations it is clear that the slit projected area on to the comet also plays 
an important role in deciding the G/R 
ratio. This point can be better understood from Table~\ref{tab-gdist} where the G/R
ratio is presented for a projected square slit on the comet at different geocentric distances. 
It is clear from this 
table that for a given physical condition of a comet and at a  given heliocentric distance, the observed 
G/R ratio for a given slit size can vary according to the geocentric distance of the 
comet. For example, for a O($^1$S) yield of 0.2\% (0.5\%)  and CO$_2$ abundance of 1\%, 
the G/R ratio can be 0.17 (0.26) if the comet is very close to the Earth (0.1 AU), whereas 
the G/R ratio can be 0.07 (0.1), 0.06 (0.08), or 0.06 (0.07), if the comet, at the time 
of observation, is at a larger distance of 0.5, 1, and 2 AU from the Earth, respectively.
Further, a G/R ratio of $\sim$0.1 can be obtained even for the  O($^1$S) yield of 0\%.
This suggests that the value of 0.1 for the G/R ratio is in no way a definitive benchmark
value to conclude that H$_2$O is the parent of atomic oxygen atom in the comet,
  since smaller ($\sim$5\% relative to H$_2$O) amounts of CO$_2$ and CO itself can produce 
enough O($^1$S) compared to that from H$_2$O. 
This table also shows that for observations made around a geocentric distance of 1 AU, the   
G/R  ratio would be generally closer to 0.1. The G/R ratio observed in different comets ranges 
from 0.02 to 0.3 \citep[e.g.,][]{Cochran08,Capria10}.

 Thus, we can conclude that the G/R  
ratio not only depends on the production and loss mechanisms of O($^1$S) atom, but also depends on the 
nucleocentric slit projected area over the comet. Moreover, the  CO$_2$ plays an important role in the 
production of O($^1$S), and thus the green line emission, in comets. With the present model calculations 
and based on the literature survey of dissociation 
channels of H$_2$O, we suggest that the O($^1$S) yield from photodissociation of H$_2$O cannot be 
more than 1\% of the total absorption cross section of H$_2$O at solar Ly-$\alpha$ radiation.
 The best fit value of O($^1$S) 
yield derived from Figure~\ref{o1so1d-cmp} for a smaller (1\%) CO$_2$ abundance
in the comet \hyak\ is 0.4 ($\pm$0.1)\%. As per the  Tables~\ref{tab-prlos1s} and~\ref{tab-prlos1d},
this means  that the ratio of rates of O($^1$S) to O($^1$D) production 
in the H$_2$O photodissociation should be 
0.03 ($\pm$0.01), which is much smaller than the value of 0.1 generally used in literature based on
\cite{Festou81}. Further, if the source of red and green lines is CO$_2$ (CO), the ratio of 
photorates for O($^1$S) to O($^1$D) would be around 0.6 (0.8) 
(see Tables ~\ref{tab-prlos1s} and~\ref{tab-prlos1d}).

To verify whether the O($^1$S) yield of 0.5\% (for the CO$_2$ abundance of 1\%) derived 
from Figure~\ref{o1so1d-cmp}, based on the comparison  between model and observed red and green 
line radial profiles in comet Hyakutake on 1996 March 30, is consistent with the G/R
 ratio observed on other 
days on this comet, we present in Table~\ref{tab-days} the G/R ratio calculated 
for observations made on 1996 March 9, 23, 27, and 30,  along with the observed value  of G/R
 ratio from \cite{Morrison97} and \cite{Cochran08}. These calculations are 
made by taking the solar flux on the day of observation using \cite{Tobiska04} SOLAR2000 model
and scaled according to the heliocentric distance  of the comet on that date. The CO abundance 
is 22\%, same as in all the calculations presented in the paper.

The calculated G/R ratio on March 9, when geocentric distance was 0.55 AU and 
 H$_2$O production rate  5 $\times$ 10$^{28}$ s$^{-1}$, is 0.09 (see Table~\ref{tab-days}) 
which is same as the observed ratio obtained by 
\cite{Cochran08}. On March 23 and 27 the comet is closer to both Sun and Earth (geocentric distance
$\sim$0.1 AU) and its H$_2$O production rate was 4 times higher than the value on March 9. 
The calculated G/R ratio on March 23 is 0.12, which is in agreement  with the observed ratio
 obtained by \cite{Morrison97}.

\section{Conclusions}
The Green and red-doublet atomic oxygen emissions are observed in comet \hyak\ in 1996 March 
when it was passing quite close to the Earth ($\Delta$ = 0.1 to 0.55 AU). 
A coupled chemistry-emission model has been developed to study the production of 
 green (5577 \AA) and red-doublet (6300 and 6364 \AA) emissions in comets. 
This model has been applied to comet Hyakutake and the results are compared with the 
observed radial profiles of 5577 and 6300 \AA\ line emissions and the green to red-doublet intensity
ratio. The important results from the present model calculations can be summarized as following.
It may be noted that some of these results enumerated below  may vary for other comets 
having different gas production rate or heliocentric distance.

\begin{enumerate}
\item The photodissociation of H$_2$O  is the dominant production process for the formation 
of O($^1$D) throughout the inner cometary coma.
  The solar H Ly-$\alpha$ (1216 \AA) flux  mainly governs  the production of O($^1$D)
in the photodissociative excitation of H$_2$O, but near the nucleus  solar radiation in 
the wavelength band 1375--1165 \AA\ can control the formation of O($^1$D) from H$_2$O. 

\item Other than the photodissociation of H$_2$O molecule, above cometocentric distance of 100 km
 the radiative decay of O($^1$S) to O($^1$D) (via 5577 \AA\ line emission), while above 1000 km 
the dissociative recombination of H$_2$O$^+$ ions,  are also significant source mechanisms
 for the  formation of O($^1$D) and O($^1$S) atoms.

\item The collisional quenching of O($^1$D) atoms  by H$_2$O is significant up to radial distance
 of  $\sim$1000 km; 
above this  distance the radiative decay is the main loss mechanism
 of O($^1$D) atoms. The collisional quenching of  O($^1$D) by other neutral species is an 
order of magnitude smaller.

\item The photodissociation of H$_2$O  is the major process for the production of O($^1$S) atoms, but
near the nucleus the photodissociation of CO$_2$ can be the dominant source.
 The solar H Ly-$\alpha$ (1216 \AA) flux controls  the production of O($^1$S) 
via photodissociative excitation of H$_2$O.

\item At cometocentric distances of $<$100 km, the main loss process for  O($^1$S)  
 is quenching by H$_2$O  molecule, while above 100 km the radiative decay 
is the dominant loss process.

\item Since the photoabsorption cross section of CO$_2$ molecule is quite small at 1216 \AA,
 the contribution of CO$_2$ in the production of O($^1$S) and O($^1$D) at the
solar H Ly-$\alpha$ is insignificant.

\item Because the CO$_2$ absorption cross section in the 1165--955 \AA\  wavelength range  
is higher by an order of magnitude compared to that at other wavelengths,  
the solar radiation in this wavelength region mainly controls the
 production of O($^1$D) and O($^1$S) in the photodissociative excitation of CO$_2$.
 Moreover,  the 
CO$_2$ absorption cross section in this band is also the largest compared to those of H$_2$O
and CO.

\item The cross section for the photodissociation of H$_2$O  producing 
 O($^1$S) at the solar H Ly-$\alpha$ wavelength (with 1\% O($^1$S) yield) is smaller
by more than two orders of magnitude than the cross section for the photodissociation of CO$_2$ producing
O($^1$S) in the wavelength region 1165--955 \AA\ . Though the solar flux at 1216 \AA\ is higher  
compared to that  in  the 1165--955 \AA\ wavelength region by two orders of magnitude, the larger value
 of CO$_2$ cross section in this wavelength band
enables CO$_2$ to be an important source for the production 
of metastable O($^1$S) atom.

\item In the case of CO, the dissociation and ionization thresholds are close to each other.
 Hence, most of the solar radiation ionizes CO molecule rather than producing the 
O($^1$S) and O($^1$D) atoms.

\item Though the CO abundance is relatively high ($\sim$22\%) in comet \hyak, the contribution of CO
photodissociation in the O($^1$D) production  is small ($\sim$1\%), while for the production of O($^1$S) 
its contribution  is 10 to 25\%. 

 \item The photoelectron impact dissociative excitation of H$_2$O, CO$_2$, and CO makes 
only a minor contribution ($<$1\%) in the formation of metastable O($^1$S) and O($^1$D) 
atoms in the inner coma.

\item The O($^1$S) density peaks at shorter radial distances than the O($^1$D) density. The 
peak value of O($^1$S) density is found around 60 km from the nucleus, while for the O($^1$D)
a broad peak around 200-600 km is observed.

\item In a H$_2$O-dominated comet, the green line emission is mainly generated in the 
 photodissociative excitation of H$_2$O with contribution of 40 to 60\% (varying according to the radial
distance) to the total  intensity, while the photodissociation of 
 CO$_2$ is the next potential source contributing 10 to 40\%.

\item For the red line emission the major source  is 
photodissociative excitation of H$_2$O,  with contribution varying from 60 to 90\% depending on 
 the radial distance from the nucleus.

\item The G/R ratio depends not only on the production and loss processes of the O($^1$S) and O($^1$D)
atoms, but also on the size of observing slit and the geocentric distance of comet at the time of
 observation. 

\item For a fixed slit size, the calculated value of the G/R ratio is found to vary between 0.03 and 0.5 depending on the
 geocentric distance of the comet.
 In the inner  ($<$300 km) most part of the coma, the G/R ratio is always larger than 0.1, with 
values as high as 0.5. On the other hand, at cometocentric distances larger than 1000 km the G/R ratio 
is always less than 0.1.

\item The model calculated radial profiles of 6300 and 5577 \AA\ lines are consistent with the 
observed profiles on comet \hyak\ for O($^1$S) yield of 0.4 ($\pm$0.1) and CO$_2$ abundances of 1\%.

\item The  model calculated G/R ratio on comet Hyakutake is in good agreement with the G/R 
ratio observed on two  days  in 1996 March by two  observatories using different slit sizes.
\end{enumerate}

 \section*{Acknowledgments}
 S. Raghuram was supported by the ISRO Senior Research Fellowship during the period of this work.

\clearpage

\begin{deluxetable}{lllll}  
\tablecolumns{3}
\tablewidth{0pc}
 
\tablecaption{Reactions for the production and loss of O($^1$S)\label{tab-prlos1s}.}
\tablehead{
\colhead{Reaction} & \colhead{Rate (cm$^{3}$ s$^{-1}$ or s$^{-1}$)}&
\colhead{ Reference}
}
\startdata
 
H$_2$O + h$\nu$ $\rightarrow$ O($^1$S) + H$_2$ 
& 6.4 $\times$ 10$^{-8}$ \tablenotemark{a} & This work \nl  
OH + h$\nu$  $\rightarrow$ O($^1$S) + H
&6.7$\times$ 10$^{-8}$  &\cite{Huebner92}\nl  
CO$_2$  + h$\nu$ $\rightarrow$ O($^1$S) + CO
& 7.2 $\times$ 10$^{-7}$ & This work\nl 
CO  + h$\nu$ $\rightarrow$  O($^1$S) + C
& 4.0 $\times$ 10$^{-8}$ & \cite{Huebner79}  \nl
H$_2$O + e$_{ph}$ $\rightarrow$ O($^1$S)  + others
& 9.0 $\times$ 10$^{-10}$ & This work \nl  
OH + e$_{ph}$ $\rightarrow$ O($^1$S) + others
& 2.2 $\times$ 10$^{-10}$ & This work \nl  
CO$_2$ + e$_{ph}$ $\rightarrow$ O($^1$S) + others
&4.4 $\times$ 10$^{-8}$ & This work\nl  
CO + e$_{ph}$ $\rightarrow$ O($^1$S) + others
& 2.2 $\times$ 10$^{-10}$ & This work \nl  
O + e$_{ph}$ $\rightarrow$ O($^1$S) 
& 3.0 $\times$ 10$^{-8}$ &This work\nl  
H$_2$O$^+$ + e$_{th}$ $\rightarrow$ O($^1$S) + others
& 4.3 $\times$ 10$^{-7}$ (300/T$_e$)$^{0.5}$ $\times$ 0.045 \tablenotemark{b} 
& \cite{Rosen00}\nl
OH$^+$ + e$_{th}$ $\rightarrow$ O($^1$S) + others
&6.3 $\times$ 10$^{-9}$ $\times$ (300/T$_e$)$^{0.5}$ & \cite{Gueberman95} \nl  
CO$_2^+$ + e$_{th}$ $\rightarrow$ O($^1$S) + others
&2.9 $\times$ 10$^{-7}$ $\times$ (300/T$_e$)$^{0.5}$&  \cite{Mitchell90} \nl  
CO$^+$ + e$_{th}$ $\rightarrow$ O($^1$S) + others
&5.0 $\times$ 10$^{-8}$ $\times$ (300/T$_e$)$^{0.46}$ & \cite{Mitchell90} \nl   
O($^1$S) + h$\nu$ $\rightarrow$ O$^+$ + e
& 1.9 $\times$ 10$^{-7}$  &\cite{Huebner92} \nl  
O($^1$S) + e$_{ph}$ $\rightarrow$ O$^+$ + 2e
& 2.7 $\times$ 10$^{-7}$ & This work \nl   
O($^1$S) \hspace{0.5cm} $\longrightarrow$ O($^3$P) + h$\nu_{2972}$
& 0.075  & \cite{Wiese96} \nl  
O($^1$S) \hspace{0.5cm} $\longrightarrow$ O($^1$D) + h$\nu_{5577}$
& 1.26  & \cite{Wiese96}\nl   
O($^1$S) + H$_2$O $\rightarrow$ 2 OH
& 3 $\times$ 10$^{-10}$ & \cite{Zipf69}\nl  
\hspace{2.4cm} $\rightarrow$ O($^1$D) + H$_2$O
& 3 $\times$ 10$^{-10}$ $\times$ 0.01 \tablenotemark{c}  & \cite{Zipf69} \nl  
O($^1$S) + CO$_2$ $\rightarrow$ O($^3$P) + CO$_2$
& 3.1 $\times$ 10$^{-11}$ exp(-1330/T)& \cite{Atkinson72} \nl
\hspace{2.4cm} $\rightarrow$ O($^1$D) + CO$_2$  
& 2.0 $\times$ 10$^{-11}$ exp({-1327/T}) 
&\cite{Capetanakis93}\nl  
O($^1$S) + CO $\rightarrow$ CO + O 
&3.21 $\times$ 10$^{-12}$ exp(-1327/T) &\cite{Capetanakis93}\nl
\hspace{2.3cm} $\rightarrow$ O($^1$D) + CO & 7.4 $\times$ 10$^{-14}$ exp({-961/T})
 & \cite{Capetanakis93}\nl 
O($^1$S) + e$_{th}$ $\rightarrow$  O($^1$D)  + e
& 8.56 $\times$ 10$^{-9}$  & \cite{Berrington81}\nl  
\hspace{2.1cm} $\rightarrow$  O($^3$P)  + e & 1.56 $\times$ 10$^{-9}$ (T$_e$/300)$^{0.94}$
& \cite{Berrington81}\nl  
O($^1$S) + O $\rightarrow$ 2 O($^1$D) &
2.0 $\times$ 10$^{-14}$  &\cite{Krauss75}\nl  
\hline
\enddata

\tablenotetext{a}{This rate is calculated assuming 1\%  yield for the production of
 O($^1$S) at 1216 \AA\ wavelength.} 
\tablenotetext{b}{0.045 is the assumed branching ratio for the formation of O($^1$S) 
via dissociative recombination of H$_2$O$^+$ ion.}
\tablenotetext{c}{0.01 is the assumed yield for the formation of O($^1$D) via quenching of H$_2$O}
\tablecomments{The  photorates and photoelectron impact rates are at 1 AU on 1996 March 30; 
e$_{ph}$ = photoelectron, e$_{th}$ = thermal electron, h$\nu$ = solar photon, 
T$_e$ = electron temperature, T = neutral temperature.}

\end{deluxetable}

\begin{deluxetable}{lllll} 
\tablecolumns{3}
\tablewidth{0pc}
\tablecaption{Reactions for the production and loss of O($^1$D)\label{tab-prlos1d}.}
\tablehead{
\colhead{Reaction} & \colhead{Rate (cm$^{3}$ s$^{-1}$ or s$^{-1}$)}&
\colhead{ Reference}
}
\startdata  
H$_2$O + h$\nu$ $\rightarrow$ O($^1$D) + H$_2$ & 8.0  $\times$ 10$^{-7}$& This work \\ 
OH + h$\nu$ $\rightarrow$ O($^1$D) + H 
& 6.4 $\times$ 10$^{-7}$ &\cite{Huebner92}\\ 
CO$_2$ + h$\nu$ $\rightarrow$ O($^1$D) + CO    &1.2 $\times$ 10$^{-6}$ & This work\\   
CO + h$\nu$ $\rightarrow$ O($^1$D) + C         &5.1  $\times$ 10$^{-8}$ & This work\\  
O($^1$S) \hspace{0.cm} $\rightarrow$  O($^1$D) + h$\nu_{557nm}$  
&1.26 & \cite{Wiese96} \\	 
H$_2$O + e$_{ph}$ $\rightarrow$ O($^1$D) + H$_2$ + e
& 2.1 $\times$ 10$^{-10}$ & This work \\ 
OH + e$_{ph}$ $\rightarrow$ O($^1$D) + H  + e
& 7 $\times$ 10$^{-11}$ & This work\\ 
CO$_2$ + e$_{ph}$ $\rightarrow$ O($^1$D) + CO + e
& 8.5 $\times$ 10$^{-9}$ & This work\\  
CO + e$_{ph}$ $\rightarrow$ O($^1$D) + C($^1$D) + e
& 7 $\times$ 10$^{-11}$ & This work\\ 
O + e$_{ph}$ $\rightarrow$ O($^1$D) & 3.7 $\times$ 10$^{-7}$  & This work \\
H$_2$O$^+$ + e$_{th}$ $\rightarrow$ O($^1$D) + H$_2$ 
&4.3 $\times$ 10$^{-7}$ (300/T$_e$)$^{0.5}$ $\times$ 0.35 \tablenotemark{a} & \cite{Rosen00} \\  
OH$^+$ + e$_{th}$ $\rightarrow$ O($^1$D) + H
& 6.3 $\times$ 10$^{-9} \times$ (300/T$_e$)$^{0.48}$ & \cite{Gueberman95}\\ 
CO$_2^+$ + e$_{th}$ $\rightarrow$ O($^1$D) + CO
&2.9 $\times$ 10$^{-7}$ (300/T$_e$)$^{0.5}$ & \cite{Mitchell90}\\ 
CO$^+$ + e$_{th}$ $\rightarrow$ O($^1$D) + C($^1$D)
&5 $\times$ 10$^{-8}$ (300/T$_e$)$^{0.46}$ &\cite{Mitchell90}\\   
O($^1$S) + e$_{th}$ $\rightarrow$ O($^1$D) + e &
1.5 $\times$ 10$^{-10}$ (T$_e/300)^{0.94}$ & \cite{Berrington81} \\  
O($^1$S) + H$_2$O $\rightarrow$ O($^1$D) + H$_2$O
& 3 $\times$ 10$^{-10}$ $\times$ 0.01 \tablenotemark{b} & \cite{Zipf69}\\ 
O($^1$S) \hspace{0.cm} + CO$_2$ $\rightarrow$  O($^1$D) + CO$_2$
&2.0 $\times$ 10$^{-11}$ exp({-1327/T}) & \cite{Capetanakis93} \\  
O($^1$S) + CO $\rightarrow$ O($^1$D) + CO & 7.4 $\times$ 10$^{-14}$ exp({-961/T}) &
\cite{Capetanakis93}\\  	 
O($^1$D) + h$\nu$  $\rightarrow$ O$^+$ + e
& 1.82 $\times$ 10$^{-7}$ & \cite{Huebner92} \\  
O($^1$D) \hspace{0.2cm} $\longrightarrow$ O($^3$P)+ h$\nu_{6300}$ 
& 6.44 $\times$ 10$^{-3}$ & \cite{Storey00}\\   
O($^1$D) \hspace{0.2cm} $\longrightarrow$ O($^3$P)+ h$\nu_{6364}$ 
&2.15 $\times$ 10$^{-3}$ & \cite{Storey00}\\  
O($^1$D) + e$_{ph}$ $\rightarrow$ O$^+$ + 2e 
& 1.75 $\times$ 10$^{-7}$ & This work \\   
O($^1$D) + e$_{th}$ $\rightarrow$ O($^3$P) + e
&8.1 $\times$ 10$^{-10}$ (T$_e$/300)$^{0.5}$ &\cite{Link82} \\   
O($^1$D) + H$_2$O  $\rightarrow$ OH + OH
& 2.1 $\times$ 10$^{-10}$ & \cite{Atkinson97}\\   
\hspace{2.5cm}  $\rightarrow$ O($^3$P) + H$_2$O
&9.0 $\times$ 10$^{-12}$ & \cite{Atkinson97}\\  
\hspace{2.5cm} $\rightarrow$ H$_2$ + O$_2$
&2.2 $\times$ 10$^{-12}$ &\cite{Atkinson97}\\   
O($^1$D) + CO$_2$ $\rightarrow$  O + CO$_2$
&7.4 $\times$ 10$^{-11}$ exp({-120/T}) & \cite{Atkinson97} \\ 
\hspace{2.5cm} $\rightarrow$ CO + O$_2$ & 2.0 $\times$ 10$^{-10}$ & \cite{Atkinson97}\\  
O($^1$D) + CO  $\rightarrow$  O + CO 
&5.5 $\times$ 10$^{-10}$ exp({-625/T}) & \cite{Schmidt88} \\   
\hspace{2.5cm} $\rightarrow$  CO$_2$
& 8.0 $\times$ 10$^{-11}$ & \cite{Demore97}\\   
\hline \enddata
\tablenotetext{a}{0.35 is the assumed branching ratio for the formation of O($^1$D) via 
dissociative recombination of H$_2$O$^+$ ion.}
\tablenotetext{b}{0.01 is the assumed branching ratio for the formation of O($^1$D) via 
quenching of H$_2$O.}
 \tablecomments{The  photorates and photoelectron impact rates are at 1 AU on 1996 March 30,
e$_{ph}$ = photoelectron, e$_{th}$ = thermal electron, h$\nu$ = solar photon, 
T$_e$ = electron temperature, T = neutral temperature.}
\end{deluxetable}
\clearpage

\newpage
\pagestyle{plain}
\begin{deluxetable}{|c|c|c|c|c|c|c|c|c|c|c|c|c|c|c|c|c|c|c|} 
\tablecolumns{19} 
\tablewidth{0pc}
\tabletypesize
\scriptsize
\rotate
\tablecaption{\small Calculated percentage contribution for major production processes of 
O($^1$S) and O($^1$D)   atoms in comet \hyak\ with varying relative abundance of 
 CO$_2$ and O($^1$S) yield. \label{tabprj-yld}} 
\tablehead{ 
{O($^1$S)} &  
\multicolumn{18}{|c|}{ Production processes of O($^1$S) and O($^1$D) at three
 cometocentric projected distances (km)}   \\  [-1 pt]
 \cline{2-19} 
 Yield\tablenotemark{a} &\multicolumn{3}{c|}{ h$\nu$ + H$_2$O} 
&\multicolumn{3}{c|}{ h$\nu$ + OH} 
&\multicolumn{3}{c|}{ h$\nu$ + CO$_2$}
&\multicolumn{3}{c|}{ e + H$_2$O$^+$} 
&\multicolumn{3}{c|}{ O($^1$S) $\rightarrow$  O($^1$D)} 
&\multicolumn{3}{c|}{ h$\nu$ + CO } \\  
 \cline{2-19}
(\%)&10$^2$&10$^3$&10$^4$&10$^2$&10$^3$&10$^4$&10$^2$&10$^3$&10$^4$&10$^2$&10$^3$&10$^4$
 &10$^2$&10$^3$&10$^4$&10$^2$&10$^3$&10$^4$\\  [-10 pt]}
\startdata 
\bf{1\% CO$_2$}&&&&&&& && & &  & & & &  & &  &     \\ [-0 pt] 
0.0 & 0 [92]\tablenotemark{b} & 0 [82] & 0 [61] & 1 [0.5] & 5 [1] & 19 [8]& 39 [1] & 23 [1]& 
      14 [1]  &  10 [2] & 28 [9] & 27 [12] & [1] & [3] & [4] &  37 [1] & 24 [1] & 16 [1] \\   
0.2 & 38 [91] & 28 [81] &  17 [60] & 1 [0.5] & 4 [1] & 15 [8]& 24 [1] & 17 [1]& 
      12 [1]  & 6 [2] & 20  [9] & 22 [12]  & [3] & [4) & [5] & 23 [1] & 17 [1] & 13  [1]\\  
0.5 & 62 [90] & 50 [80] & 34 [60] & 0.5 [0.5] & 3 [1] & 12 [8] & 15 [1] & 11 [1]& 
      10 [1]  &  4 [2] & 14 [9] & 17 [12] & [4] & [6] & [6] & 15 [1] & 12 [1] & 10 [1]  \\  
1.0 & 75 [88] & 66 [77] & 51  [58] & 0.5 [0.5] & 2 [1] & 10 [7]& 9 [1] & 8 [1]& 
      7 [1]  & 2 [2] & 10 [9] & 13 [12] & [7] & [9] & [9] & 9 [1] & 8 [1] & 7 [1]\\ [-0pt]
\bf{0\% CO$_2$}&&&&&&& && & &  & & & &  & &  &    \\  [-0pt]

0.0 &  0 [95] & 0 [84] & 0 [62] &  2 [0.5] & 7 [2] &  23  [8]& 0 [0] &  0 [0]& 
      0 [0]  &  17 [2] & 39 [9] & 34 [12] & [1] & [2] & [3] & 65 [1] & 34 [1] & 20 [1] \\ 
0.2 & 51 [94] & 35 [83] & 21 [62] & 1 [0.5] & 5 [1] & 18 [8]&  0 [0] & 0 [0]& 
        0 [0 ]  & 8 [2]   & 25 [9] & 27 [12] & [2] & [3] & [4] &  31 [1] & 22 [1] & 16 [1] \\  
0.5 & 72 [92] & 57 [81] & 40 [61] & 0.5 [0.5] & 3 [1] & 14 [8] & 0 [0] &  0 [0]& 
       0 [0]  & 5 [2]   & 16 [9] & 20 [12] & [4] & [5] & [5] & 17 [1] & 14 [1] & 12 [1]\\  
1.0 & 84 [90] & 73 [79] & 57 [60] & 0.5 [0.5] & 2 [2] & 10 [8]&  0 [0] & 0 [0]& 
        0 [0] & 2 [2]   & 10 [9] & 14 [12]   & [6] & [8] & [8] & 10 [1] & 9 [1] & 8  [1]\\  [-0pt]
\bf{3\% CO$_2$}&&&&&&& && & &  & & & &  & &  &    \\  [-0pt]
0.0 & 0 [89] & 0 [79] & 0 [58] & 1 [0.5] & 3 [1] & 13 [7]& 62 [4] & 44 [4]& 
      30 [3]  &  5 [2] & 18 [9] & 19 [12] & [3] & [4] & [5] &  20 [1] & 15 [1] & 11 [1] \\  [-0pt]
\bf{5\% CO$_2$}&&&&&&& && & &  & & & &  & &  &    \\  [-0pt]
0.5 &  36 [83] & 31 [72] & 22 [54] &  0.5 [0.5] & 2 [2] & 7 [8]& 45 [1] & 37 [1]& 
      30 [1]  & 2 [2] & 10 [10] & 12 [12] & [7] & [8] & [9] & 9 [1] & 8 [1] & 7 [1] \\  
\enddata 

\tablenotetext{a}{ \tiny Yield for the production of O($^1$S) from photodissociation of H$_2$O at 
solar Lyman-$\alpha$ (1216 \AA) line.} 
\tablenotetext{b}{\tiny  The values in square brackets are for the  O($^1$D).} 
\tablecomments{\tiny Calculations are made for 
1996 March  30, when r = 0.94  AU and $\Delta$ = 0.19 AU.}
\end{deluxetable} 

\begin{deluxetable}{|c|c|c|c|c|c|c|c|c|c|c|c|c|c|c|c|c|c|c|c|c|c|c|} 
\tablecolumns{8} 
\tablewidth{0pc} 
\tabletypesize
\scriptsize
\rotate
\tablecaption{Calculated percentage contribution for the major production processes
 of the green (red-doublet) emission in the slit projected field of view on comet \hyak. 
\label{tabprj-slit}}
\tablehead{
 \multicolumn{1}{c|}{ O($^1$S) Yield  (\%)} 
&\multicolumn{1}{c|}{ h$\nu$ + H$_2$O} 
&\multicolumn{1}{c|}{ h$\nu$ + OH}
&\multicolumn{1}{c|}{ h$\nu$ + CO$_2$}
&\multicolumn{1}{c|}{ e$^-$ + H$_2$O$^+$} 
&\multicolumn{1}{c|}{ O($^1$S) $\rightarrow$ O($^1$D)} 
&\multicolumn{1}{c|} {\centering h$\nu$ + CO} 
&\multicolumn{1}{c|}{ G/R ratio\tablenotemark{b}} \\ [-10pt]
}
\startdata
\bf{1\% CO$_2$}   & &&& &  & &   \\ [-0pt]

 0.0 & 0 [91]\tablenotemark{a} & 2 [0.5]& 36 [1] & 13 [3] &  [1] & 35 [1]& 0.07  \\  
 0.2 & 36 [91] & 1 [0.5] & 23 [1] & 8 [3] &  [3]& 22 [1]& 0.11  \\
 0.5 & 59 [89] & 1 [0.5] & 14 [1] & 5 [3] &  [4] & 14 [1]& 0.17  \\      
 1.0 & 76 [87] & 0.5 [0.5]& 10 [1] & 0.5 [3] &  [6]& 10 [1]& 0.27   \\   
           
\bf{0\% CO$_2$}  & &&& &  & & \\
 0.0 & 0 [94] & 4 [0.5]& 0 [0] &  21 [3] &  [1]&  59 [1] & 0.04 \\   
 0.2 & 49 [93] & 2 [0.5] &0 [0] & 11 [3] & [2] &  30 [1] & 0.08 \\
 0.5 & 70 [91] & 1 [0.5] &0 [0] & 6 [3] & [4] &   17 [1] & 0.15  \\  
 1.0 & 82 [89] & 0.5 [0.5] &0 [0] & 3 [3] & [6] & 10 [1] & 0.25   \\    
\bf{3\% CO$_2$}  & &&& &  & & \\
 0.0 & 0 [87] & 1 [0.5] & 60 [4] & 7 [3] & [3] & 20 [1] & 0.13 \\           
\bf{5\% CO$_2$}  & &&& &  & & \\  
0.5 & 35 [82] & 0.5 [0.5] & 45 [6] &  3 [3] & [7] &  7 [1] & 0.27 \\   

\enddata
\tablenotetext{a}{The values in square brackets are the calculated percentage  contribution for 
the red-doublet emission.} 
\tablenotetext{b}{The calculated values are averaged over the projected 
area of 165 $\times$ 1130 km corresponding to slit size of 1.2$''$ $\times$ 8.2$''$ 
 at $\Delta$ = 0.19 AU  centred on the nucleus of comet \hyak\  
on 1996 March 30  \citep{Cochran08}.}
\end{deluxetable} 

\begin{deluxetable}{|c|c|c|c|c|c|c|c|c|c|c|c|c|c|c|c|c|c|c|c|c|c|c|} 
 \tablecolumns{8} 
\tablewidth{0pc} 
\tabletypesize
 \scriptsize
\tablecaption{Calculated green to red-doublet emission brightness 
ratio averaged over 5$''$ $\times$ 5$''$ slit, 
at different geocentric distances.\label{tab-gdist}}
\rotate
\tablehead{
\multicolumn{1}{c}{{\centering Yield\tablenotemark{a}}  } &
 \multicolumn{6}{c|}{Geocentric distance (AU)}\\ 
\cline{2-7}
(\%)&\multicolumn{1}{c|}{0.1}
&\multicolumn{1}{c|} {0.2}
&\multicolumn{1}{c|} {0.5}
&\multicolumn{1}{c|} {1}        
&\multicolumn{1}{c|} {1.5}
&\multicolumn{1}{c|} {2}  \\ [-10pt]
}
\startdata
\bf{1\% CO$_2$}   &&&  &&& \\
0.0  & 0.11 & 0.07  & 0.05 & 0.04 & 0.04 & 0.04 \\
0.2  & 0.17 & 0.11  & 0.07 & 0.06 & 0.05 &  0.05 \\
0.5  & 0.26 & 0.17  & 0.10 & 0.08 & 0.07 & 0.07 \\
1.0  & 0.40 & 0.26  & 0.15 & 0.12 & 0.10 & 0.10 \\

\bf{0\% CO$_2$}   &&&  &&& \\
0.0  & 0.07  & 0.05  & 0.03  & 0.03 & 0.03  & 0.03 \\
0.2  & 0.13  & 0.09  & 0.05  & 0.05 & 0.04  & 0.04 \\
0.5  & 0.23  & 0.15  & 0.09  & 0.07 & 0.06  & 0.06 \\
1.0  & 0.37  & 0.24  & 0.14  & 0.11  & 0.01  & 0.01  \\

\bf{3\% CO$_2$}   &&&  &&& \\
0.0  & 0.19 & 0.13  & 0.08 & 0.06 & 0.06 & 0.06 \\
0.5  & 0.33 & 0.21  & 0.13 & 0.10 & 0.09 & 0.09 \\
\enddata
 \tablenotetext{a}{O($^1$S) yield from photodissociation of H$_2$O} 
 \tablecomments{Calculations are made for 1996 March 30, where r=0.94 AU.}
\end{deluxetable}


\begin{deluxetable}{|c|c|c|c|c|c|c|c|c|c|c|c|c|c|c|c|c|c|c|c|c|c|c|} 
\tablecolumns{8} 
\tablewidth{0pc} 
\tabletypesize
 \scriptsize
\rotate
\tablecaption{Calculated intensities of green and red-doublet emissions and the G/R ratio in 
comet \hyak\ on different days of observation in 1996 March.\label{tab-days}}
\tablehead{
\multicolumn{1}{|p{0.9 in}|}{\centering Date of observation (1996 March )}
&\multicolumn{1}{c|}{\multirow{2}{0.3 in}{\centering r }}
&\multicolumn{1}{c|}{\multirow{2}{0.3 in}{\centering $\Delta$} }
&\multicolumn{1}{c|}{\multirow{2}{0.5 in}{\centering Q$_{H_2O}$}}
&\multicolumn{1}{c|}{\multirow{2}{0.7 in}{\centering Slit dimension} }        
&\multicolumn{1}{c|} {\multirow{2}{0.7 in}{\centering Projected distance} }
&\multicolumn{1}{c|} {\multirow{2}{0.8 in}{\centering Calculated 5577 \AA\ intensity} } 
&\multicolumn{1}{c|}{\multirow{2}{0.8 in} {\centering Calculated (6300 + 6364 \AA) intensity }  }
 &\multicolumn{2}{c|}{\multirow{2}{1 in}{\centering {G/R ratio}}}  \\  
\cline{9-10}
&(AU)&(AU)&(s$^{-1}$) &(arcsec)&(km) &(kR)& (kR)&Calculated &Observed\\ [-10pt]
}
\startdata
9\tablenotemark{a} & 1.37 & 0.55 & 5 $\times$ 10$^{28}$ & 1.2$''$ $\times$ 8.2$''$& 470 $\times$ 3720 &0.06
&0.62&0.09&0.09\tablenotemark{a}\\
23\tablenotemark{b}& 1.08 & 0.12 & 1.8 $\times$ 10$^{29}$ & 7.5$''$ (circular)\tablenotemark{c} & 640 & 0.69 & 5.88 &
 0.12 &0.12 - 0.16\tablenotemark{b}\\
27\tablenotemark{b}& 1.00 & 0.11 & 2 $\times$ 10$^{29}$  & 7.5$''$ (circular) & 653 & 0.89 & 7.12 &
 0.12 &--\\
30\tablenotemark{a}& 0.94 & 0.19 & 2.2 $\times$  10$^{29}$
 & 1.2$''$ $\times$ 8.2$''$& 165 $\times$ 1129 &  0.90& 7.97& 0.17&--\\
\enddata
\tablenotetext{a}{\cite{Cochran08}}
\tablenotetext{b}{\cite{Morrison97}} 
\tablenotetext{c}{7.5$''$ is the diameter of the circular slit} 
\tablecomments{Calculations are made for O($^1$S) yield of 0.5\%, and CO$_2$ and CO 
relative abundances of 1\% and 22\%, respectively.}

\end{deluxetable}

\clearpage
\newpage

\clearpage

\end{document}